\documentclass[usaAMS,usenatbib]{mn2e}
\usepackage{amsmath,amssymb}
\usepackage{graphicx,epsfig, color}
\usepackage{natbib}
\usepackage{psfrag}
\usepackage{rotating}
\usepackage[abs]{overpic}
\usepackage{comment}
\usepackage{afterpage}
\date{\today}
\bibpunct{(}{)}{;}{a}{}{,}
\newcommand {\apgt} {\ {\raise-.5ex\hbox{$\buildrel>\over\sim$}}\ }
\newcommand {\aplt} {\ {\raise-.5ex\hbox{$\buildrel<\over\sim$}}\ }

\def\ie{{\rm i.e.}}

\def\spose#1{\hbox to 0pt{#1\hss}}
\def\ltsimm{\mathrel{\spose{\lower 3pt\hbox{$\sim$}}
        \raise 2.0pt\hbox{$<$}}}
\def\gtsimm{\mathrel{\spose{\lower 3pt\hbox{$\sim$}}
        \raise 2.0pt\hbox{$>$}}}

\def\cm{{\rm\thinspace cm}}

\def\s{{\rm\thinspace s}}

\def\g{{\rm\thinspace g}}

\def\erg{{\rm\thinspace erg}}

\def\Hz{{\rm\thinspace Hz}}

\def\ster{{\rm\thinspace ster}}
\def\ergps{\hbox{${\rm\erg\s^{-1}\,}$}}

\def\pcm{\hbox{${\rm\cm^{-1}\,}$}}
\def\pcm2{\hbox{${\rm\cm^{-2}\,}$}}
\def\pcm3{\hbox{${\rm\cm^{-3}\,}$}}
\def\ergpscm3Hz{\hbox{${\rm\ergps\cm^{-3}\Hz^{-1}\,}$}}
\def\ergpscm3Hzster{\hbox{${\rm\ergps\cm^{-3}\Hz^{-1}\ster^{-1}\,}$}}
\def\gpcm3{\hbox{${\rm\g\cm^{-3}\,}$}}
\def\ergpcm2{\hbox{${\rm\erg\cm^{-2}\,}$}}
\def\ergpcm3{\hbox{${\rm\erg\cm^{-3}\,}$}}
\def\phpscm2{\hbox{${\rm photons\s^{-1}\cm^{-2}\,}$}}

\def\Aluzas{Al\={u}zas}

\title[Shocks and magnetized clumpy regions]{Numerical simulations of
  a shock interacting with multiple magnetized clouds}
\author[R.\Aluzas ~et al.]
{R. \Aluzas$^{1}$\thanks{E-mail: js07ra@leeds.ac.uk}, J.~M.~Pittard$^{1}$, S.~A.~E.~G.~Falle$^{2}$, T.~W.~Hartquist$^{1}$\\
$^{1}$School of Physics and Astronomy, University of
       Leeds, Woodhouse Lane, Leeds LS2 9JT, UK\\
$^{2}$Department of Applied Mathematics, University of 
	Leeds, Woodhouse Lane, Leeds LS2 9JT, UK
}
\begin{document}

\date{Accepted ... Received ...; in original form ...}

\maketitle

\begin{abstract}
  We present  2D adiabatic magnetohydrodynamic (MHD)  simulations of a
  shock interacting  with groups of  two or three  cylindrical clouds.
  We study how the presence  of a nearby cloud influences the dynamics
  of  this  interaction, and  explore  the  resulting differences  and
  similarities  in  the evolution  of  each  cloud. The  understanding
  gained  from  this small-scale  study  will  help  to interpret  the
  behaviour of systems with many 10's or 100's of clouds.

  We observe a wide variety  of behaviour in the interactions studied,
  which is  dependent on the initial  positions of the  clouds and the
  orientation and  strength of the  magnetic field.  We find:  i) some
  clouds  are stretched  along their  field-lines, whereas  others are
  confined by  their field-lines;  ii) upstream clouds  may accelerate
  past downstream  clouds (though magnetic tension  can prevent this);
  iii) clouds  may also change their relative  positions transverse to
  the direction  of shock propagation as they  ``slingshot'' past each
  other; iv) downstream clouds may be offered some protection from the
  oncoming flow as a result of  being in the lee of an upstream cloud;
  v)  the cycle  of cloud  compression and  re-expansion  is generally
  weaker  when there are  nearby neighbouring  clouds; vi)  the plasma
  $\beta$   in   cloud   material   can   vary   rapidly   as   clouds
  collide with one another, but low values of $\beta$ are always transitory.

  This work is relevant to studies of multi-phase regions, where fast,
  low-density gas interacts with dense clouds, such as in
  circumstellar bubbles, supernova remnants, superbubbles and
  galactic winds.

\end{abstract}


\section{Introduction}
The interstellar medium (ISM) is  recognized to be highly dynamic.  At
any  given  time  a  substantial  quantity  of  gas  is  found  to  be
transitting    between   several    different   phases    of   thermal
equilibrium. Such transitions  are driven by a variety  of heating and
cooling mechanisms. Heating is dominated by vigorous energy input from
high-mass  stars, including their  intense ionizing  radiation fields,
their powerful winds, and their terminal supernova explosions. Heating
also occurs  via the conversion of gravitational  potential energy and
from the impact  of extragalactic material. Cooling is  achieved via a
multitude of radiative processes and through adiabatic expansion.

Given  these  conditions, it  is  not  uncommon  for hot,  high  speed
material to interact with cooler, dense material (often referred to as
clouds). Knowledge  of the dynamical  and thermal behaviour of  gas in
such  interactions is necessary  for a  complete understanding  of the
nature of  the ISM.  For instance, in  starburst galaxies,  the energy
input from  high-mass stars inflates superbubbles which  can burst out
of their host. However, the properties of such flows may be controlled
by their interaction with small clouds which dominate the mass in such
regions.  These clouds may  be destroyed  and their  mass incorporated
into  the  hot  phase,  a  process  known  as  ``mass-loading''.  This
behaviour  is a  key  ingredient  in models  of  galaxy formation  and
evolution  \citep[e.g.,][]{2010MNRAS.409.1541S}, but is  currently not
calculated  self-consistently   in  them.  On  the   other  hand,  the
compression  of clouds  by the  flow may  ultimately trigger  new star
formation.

By far  the best studied case is  that of a shock  hitting an isolated
spherical  cloud.  The  hydrodynamics  of the  interaction  have  been
reported in  a number of papers  in which the  cloud density contrast,
$\chi$,  and the  shock  Mach  number, $M$,  have  been varied  (e.g.,
\citealt{1992ApJ...390L..17S},           \citealt{1994ApJ...420..213K},
\citealt{2006ApJS..164..477N}). The effect  of other processes in this
interaction  have also been  studied, such  as magnetic  fields (e.g.,
\citealt{1994ApJ...433..757M},          \citealt{2008ApJ...680..336S}),
radiative      cooling      (e.g.,      \citealt{2002A&A...395L..13M},
\citealt{2004ApJ...604...74F},   \citealt{2010ApJ...722..412Y}),   and
thermal      conduction      (e.g.,     \citealt{2005A&A...444..505O},
\citealt{2008ApJ...678..274O}).   The    turbulent   nature   of   the
destruction   of    clouds   has   been    investigated   too   (e.g.,
\citealt{2009MNRAS.394.1351P}, \citealt{2010MNRAS.405..821P}).  In the
purely  hydrodynamic  case clouds  are  destroyed  via  the growth  of
Kelvin-Helmholtz  (KH)  and  Rayleigh-Taylor (RT)  instabilities.  The
interaction becomes milder at lower  shock Mach numbers, with the most
marked differences  occuring when the post-shock gas  is subsonic with
respect to  the cloud. Cloud  density contrasts $\chi  \gtsimm 10^{3}$
are  required for  material  stripped off  the  cloud to  form a  long
``tail-like''  feature.    Efficient  cooling  causes   the  cloud  to
fragment.

The presence  of magnetic fields can strongly  affect the interaction.
In  2D  axisymmetry, magnetic  fields  parallel  to  the shock  normal
suppress  Richtmyer-Meshkov  (RM)  and  KH instabilities,  and  reduce
mixing. The magnetic field is  amplified behind the cloud due to shock
focussing  and forms a  ``flux rope''  \citep{1994ApJ...433..757M}. In
contrast,  in  3D  simulations  with strong  fields  perpendicular  or
oblique to  the shock normal  the shocked cloud becomes  sheet-like at
late times,  and oriented parallel  to the postshock field.  The cloud
then    fragments    into    vertical   or    near-vertical    columns
\citep{2008ApJ...680..336S}. More recent work including magnetic
fields, anisotropic thermal conduction and radiative cooling of 3D
shock-cloud interactions finds that intermediate strength fields are
most effective at producing long-lasting density fragments - stronger
fields prevent compression while weak fields do not sufficiently
insulate the cloud to allow efficient cooling (e.g., \citealt{JohanssonZiegler2013}).

Relatively  few  investigations of  the  interaction  of  a flow  with
multiple clouds  exist. The response of a clumpy and magnetized
  medium to a source of high pressure was considered by
  \citealt{Elmegreen1988}, who derived jump conditions for cloud
  collision fronts under a continuum approximation. This work was extended using a multi-fluid
  formalism by \citealt{2002MNRAS.333....1W}, who showed that shocks can
  rapidly broaden and thus create a more benign environment which aids
  the survival of multiphase structure passing through the
  shock. 

Simulations in which the interaction of a flow over numerous
 obstacles is studied in detail are only just becoming feasible. However,  it is clear  that the  flow responds
differently to  the presence of a  group of clouds, with  a global bow
shock   forming  when   the   clouds  are   relatively  close   (e.g.,
\citealt{2002ApJ...576..832P},           \citealt{2005MNRAS.361.1077P},
\citealt{paper1} - hereafter Paper~I).  The degree to which the nature
of the  flow changes depends on  the relative amount of  mass added to
the  flow  by  destruction   of  the  clouds,  i.e.  the  mass-loading
factor. Simulations extending Poludnenko's work to higher mass-loading
factors were  presented by  Paper~I. This work  found that  the global
flow is not  strongly affected by the presence  of clouds with density
contrasts  of $\chi=10^2$,  as it  evolves  similarly to  a region  of
equivalent, uniform  density. However, significant  changes arise when
the cloud density contrast increases  to $\chi=10^3$. In this case the
total  mass in  the clouds  becomes dominant  at a  much  lower volume
fraction (equivalently a lower total cross-section of the clouds). The
resulting interaction does not affect the structure of the shock much,
but  significantly  mass-loads  the  post-shock  flow.   This  ongoing
mass-loading of  the flow  as the clouds  are destroyed can  cause the
shock to decelerate even after it has left the clumpy region.

The  evolution of  a cloud  also  changes when  additional clouds  are
nearby.   In isolation,  clouds lose  most  of their  mass through  KH
instabilites, with the largest scale instabilities taking some time to
grow.  In mass-loaded flows,  instabilities develop more easily due to
the turbulent nature of the flow. Clouds are also ablated more quickly
due   to   the   higher   density  of   the   mass-loaded   post-shock
flow. 

Fig.~19 in Paper~I shows that the cloud lifetimes can be reduced by as
much  as 40\%,  compared  to  the single-cloud  lifetime  at the  same
resolution.  However, we  have  since discovered  a  problem with  our
previous analysis which for computational reasons was conducted on low
resolution  single- and  multi-cloud  runs. The  problem  is that  the
development  of  KH instabilities  is  significantly  slowed at  lower
resolution and  clouds instead lose mass through  direct ablation. The
latter is a stronger effect  in the multi-cloud simulations due to the
higher  density of the  flow caused  by material  mixing into  it from
clouds further upstream.  Thus our previous low-resolution simulations
in Paper I were biased against the development of KH instabilities but
not against  direct ablation, leading us to  erroneously conclude that
clouds in multi-cloud runs have  shorter lifetimes. We now find from a
high-resolution comparison  of the lifetime  of clouds in  single- and
multi-cloud simulations  that the clouds are  destroyed in essentially
the  same time\footnote{However, the  nature of  the destruction  is a
little  different. In multi-cloud  simulations, clouds  initially lose
mass a little more slowly  than in single-cloud simulations because of
the reduction in the shock-speed  brought about by the mass-loading of
the flow.  However, as the  shocked cloud moves further  downstream it
encounters increasing post-shock  density relative to the single-cloud
case, and this increases the rate of ablation slightly. The net effect
is  that the  overall lifetime  of the  cloud is  very similar  to the
single  cloud case.  Having said  this, clouds  with a  higher density
contrast than  the majority of  neighbouring clouds \emph{do}  seem to
still   be   destroyed    more   quickly   than   their   single-cloud
counterparts.  We tentatively  suggest this  is because  of  the dense
shell of ablated material which overruns them and increases their rate
of mass-loss  from ablation (all  similar clouds are destroyed  by one
cloud destruction length ($1 L_{CD}$) behind the shock front, and so are
not affected  by the shell, whereas  the denser clouds  still exist at
the  time  they  are  overrun  by  the shell).  This  effect  will  be
investigated in a forthcoming paper.}.

MHD studies of the interaction of a shock with a single-cloud show  that the  field is  amplified not so  much in  the shear
layers and vortices but rather in regions of compression: ahead of the
cloud for  perpendicular shocks where field  lines bunch up,  and in a
``flux  rope'' behind  the  cloud  where the  flow  converges for  the
parallel-shock  case  \citep{1994ApJ...433..757M}.  These  simulations
show that magnetic  fields limit mixing and fragmentation,  but do not
stop it completely, and provide  support to the cloud perpendicular to
the field lines.  Our goal in this paper is to determine the degree to
which neighbouring  clouds change this picture. In  particular, we are
interested in the amplification of the magnetic field and the presence
of magnetically  dominated regions with $\beta<1$.  Can clouds present
in regions  of enhanced  magentic field enhance  the field  further or
does it saturate? Because of the complex nature of the interaction and
the  many free  parameters which  now also  include the  positions and
separations  of clouds, we  limit this  current study  to interactions
involving two or three clouds. For computational reasons we also limit
our study  to 2D (\ie \, our  clouds are infinite  cylinders). This work
will serve as  a basis for future work exploring  the interaction of a
shock with many 10's and 100's of clouds in 2D and 3D.

The outline of this paper is as follows. In Sec.~\ref{sec:method} we
introduce our numerical method. Sec.~\ref{sec:results} details the
results of our simulations. In 
Sec.~\ref{sec:conclusions} we summarize and conclude.

\section{Method}
\label{sec:method}

The  computations were  performed  using the  {\sc  mg} adaptive  mesh
refinement (AMR)  code. The ideal  magnetohydrodynamic (MHD) equations
are solved using  a linear Riemann solver for most  cases and an exact
solver  when  there is  a  large  difference  between the  two  states
\citep{1998MNRAS.297..265F}.  Piecewise linear  cell  interpolation is
used.  The scheme  is second order accurate in space  and time, and is
supplemented   by  a  divergence   cleaning  technique   described  in
\cite{2002JCoPh.175..645D}.

The simulations were performed on 2D $XY$-cartesian grids, so that the
clouds are  actually infinite cylinders.  Two grids ($G^0$  and $G^1$)
cover the entire domain. Finer  grids are added where they were needed
and  removed  where they  are  not.  Refinement  and derefinement  are
controlled by differences in the solutions on the coarser grids with a
tolerance of  1 per cent  in the conserved quantities  specified. Each
refinement  level increased  the  resolution in  all  directions by  a
factor of $2$. The time-step on  grid $G^n$ is $\Delta t_0 / 2n$ where
$\Delta t_0$ was the time-step  on $G^0$. Refinement is performed on a
cell-by-cell basis rather than patches.

A  typical  grid extended  $X  \in [-50:190]  \,  r_{cl}$  and $Y  \in
[-50:50]  \, r_{cl}$, where  $r_{cl}$ is  the cloud  radius (identical
clouds  are assumed).  Inflow  boundary conditions  were  used at  the
negative $X$  boundary, being set  by the shock jump  conditions. Free
inflow/outflow   conditions    were   used   at    the   other   three
boundaries. Simulations  were performed with two  sets of resolutions:
$32$  cells per  cloud radius  ($R_{32}$), and  $128$ cells  per cloud
radius ($R_{128}$).  The lower resolution  runs used $7$  grid levels,
with $\Delta  x =  2 \, r_{cl}$  on the  $G^0$ grid, while  the higher
resolution simulations  used $8$  grid levels, with  $\Delta x =  1 \,
r_{cl}$ on the $G^0$ grid.

The  simulations set  up  two or  three  clouds with  a cloud  density
contrast  of $\chi=100$  and  with soft  edges  following the  density
profile as  specified in \cite{2009MNRAS.394.1351P}  with $p_1=10$. In
all  simulations the  sonic  Mach number  of  the shock  was $3$.  The
strength of  the magnetic field and  its orientation to  the shock was
varied. Values  for the Alfv\'{e}nic Mach number,  the pre-shock field
angle  and  the plasma  $\beta$  in  different  regions are  given  in
Table~\ref{table:sims}. A  different advected scalar is  used for each
cloud to  track the cloud material.  The time is measured  in units of
the  cloud crushing  timescale, $t_{cc}  = \chi^{1/2}  r_{cl}  / v_b$,
where $v_b$ is the shock velocity in the ambient medium. The bow-shock
reaches  the  $Y$  boundaries  at  around  $7.5  \,  t_{cc}$  and  the
simulations are terminated  shortly afterwards. Adiabatic behaviour is
assumed with $\gamma= 5/3$.

\begin{table}
\caption{Summary of the magnetic field strength and orientation in the
  single- and multi-cloud simulations performed. The value of the plasma $\beta$ in the
  pre-shock (i.e. $\beta_{0}$) and post-shock regions is also provided, as well as its
  approximate value in the bow-shock region.} % title of Table
\centering % used for centering table
\begin{tabular}{c c c c c c} % centered columns (4 columns)
\hline\hline %inserts double horizontal lines
 & & & \multicolumn{3}{c}{Value of $\beta$ in each region} \\
Case name & B angle & $M_a$ & pre-shock & post-shock & bow-shock \\ [0.5ex] 
%inserts table
%heading
\hline % inserts single horizontal line
b15b1 & $15^\circ$ & 2.91 & 1.13 & 6.06 & $7.1$\\ % inserting body of the table
byb1 & $89.9^\circ$ & 2.91 & 1.13 & 1.25  & $1.2$\\
byb5 & $89.9^\circ$ & 6.16 & 5.06 & 6.05 & $5.5$\\
bxb1 & $0^\circ$ & 2.91 & 1.13 & 12.4 & $21$ \\
bxb0.5 & $0^\circ$ & 2.03 & 0.55 & 6.05 & $10.5$ \\  [1ex] %adds vertical space
%m10b12b1 & $12^\circ$ & 1.13 & ?? & ?? \\ [1ex]
\hline %inserts single line
\end{tabular}
\label{table:sims} % is used to refer this table in the text
\end{table}

\section{Results}
\label{sec:results}
The collective interactions between a large number of clouds can be
incredibly complex.  To better understand them we begin by reviewing
the basic behaviour of a shock striking an isolated, magnetized,
cylindrical cloud.  We then investigate the simplest of multiple cloud
cases, that of two clouds, before applying the insight from the
2-cloud simulations to simulations with 3 clouds.  Single-cloud
simulations are named using the format \emph{sc bAbB}, where the
``\emph{sc}'' indicates that it is of a single-cloud, the ``\emph{A}''
indicates the orientation of the field (``\emph{x}'', ``\emph{15}''
and ``\emph{y}'' indicate parallel, oblique and perpendicular shocks),
and ``\emph{B}'' indicates the value of the pre-shock plasma
$\beta$. 2-cloud simulations are named using the format \emph{s2wYoX
bAbB} (or often using the shortened forms \emph{wYoX} or \emph{wYoX
bAbB}). Similarly, 3-cloud simulations are named using the format
\emph{s3wRa$\theta$ bAbB} (again also with shortened
versions). \emph{wYoX} and \emph{wRa$\theta$} identify the relative
positions of clouds, see Sec.~\ref{sec:2cloud} and
Sec.~\ref{sec:3clouds} respectivley for further details.

\begin{figure}
  \begin{center}
    \includegraphics[width=7cm]{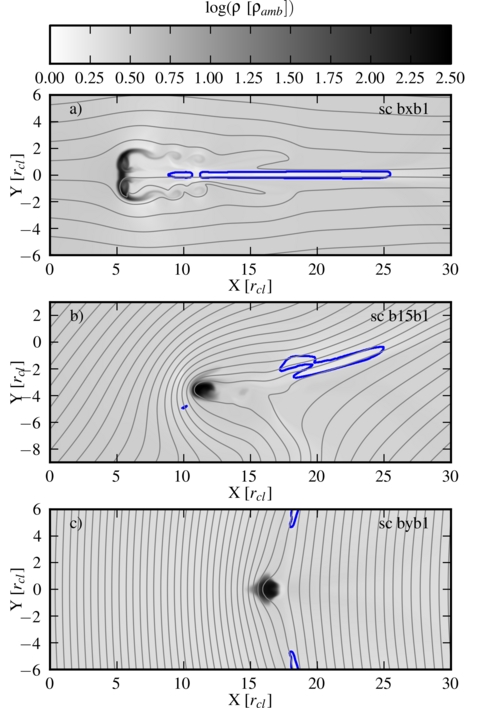}\\
    \caption{The morphology  of interactions of a shock  with a single
      cylindrical cloud.  The calculations are  in 2D, the  sonic Mach
      number is 3  and the Alfvenic Mach number  is 2.91 ($\beta_{0} =
      1.13$).  The   shock  is  a)   parallel,  b)  oblique,   and  c)
      perpendicular . The cloud  is initially positioned at the origin
      The grayscale shows the logarithmic density 
      and magnetic field lines are  also shown. 
      The  contour indicates  regions with
      low  plasma  $\beta$  and  low  momentum  ($\beta<1$  and  $\rho
      |\mathbf{u}|  < 0.5  \times  \rho_{ps} |\mathbf{u}_{ps}|$).  The
      time of the interaction is $t = 4\,t_{\rm cc}$.}
    \label{fig:singlecloud_morphology}
    \end{center}
  \end{figure}

\subsection{Single-cloud interactions}
\label{sec:1cloud}
\subsubsection{Parallel shocks}
\label{sec:1cloud_parallel}
We begin by reviewing the morphology  of the 2D interaction of a shock
with a  single magnetized, cylindrical  cloud.  In the  parallel field
case  a  ``flux-rope''  forms  directly  behind the  cloud:  the  flow
converging  behind   the  cloud  compresses  the   field  lines,  thus
increasing the  magnetic pressure  which prevents the  post-shock flow
from  entering it  (see Fig.~\ref{fig:singlecloud_morphology}a).  As a
result the  ``flux-rope'' not  only has a  low plasma $\beta$,  but it
also  has very low  momentum. These  two conditions  ($\beta <  1$ and
$\rho |\mathbf{u}| <  0.5 \times \rho_{ps} |\mathbf{u}_{ps}|$) specify
the ``flux-rope'' region  in the parallel field case,  but can also be
met in other field arrangements.

Another important  feature in  the flow are  the ``wings''. This  is a
region or regions  alongside the flux rope which  delineates where the
flow is stripping material away  from the cloud.  This region shows up
in the magnetic field structure of simulations with parallel shocks as
the reversal of the magnetic field.
In general the  ``wings'' are shielded from the  momentum of the flow,
although  occasionally  they  may  contain  higher  density  fragments
stripped off the upstream cloud.

\subsubsection{Oblique shocks}
In  our oblique  shock  simulations a  preshock  field orientation  of
$\theta_0 = 15^\circ$ was chosen  to be a representative oblique field
case.   This   gives  $\theta_{ps}  =  45^\circ$   in  the  post-shock
medium. When  an oblique shock interacts with  an isolated cylindrical
cloud we find  that the field lines wrap around  the cloud keeping its
cross-section      roughly       circular      in      shape      (see
Fig.~\ref{fig:singlecloud_morphology}b).  Field  lines above the cloud
become    nearly    parallel     to    the    direction    of    shock
propagation\footnote{The  postshock flow  is about  $-7^\circ$  to the
shock normal.}   and some material  is stripped off along  them. Field
lines  below  the  cloud span  a  range  of  angles, with  the  region
immediately upstream  of the cloud having field  lines nearly parallel
to the shock front.  Field amplification and ``shielding'' (i.e. where
gas has minimal exposure to the ambient  flow - e.g. gas in the lee of
a cloud) now occur in distinct, but overlapping regions.  The cloud is
accelerated      downstream      and      also      laterally      (in
Fig.~\ref{fig:singlecloud_morphology}b) the  cloud is seen  to move to
lower  $Y$).   The  asymmetry  of  the  cloud's  motion  reflects  the
asymmetric  bunching  and  tensioning  of  the  field  lines  and  the
direction of the postshock flow.  Note that because the cloud in
  this simulation is actually an infinite cylinder field lines cannot
  easily slip past it. If the cloud were spherical we would expect
  some splitting and rearranging of the field, which could
  significantly change the forces acting on the cloud.

\subsubsection{Perpendicular shocks}
In the perpendicular field case, the magnetic field is initially
amplified directly upstream of the cloud where the flow stagnates
against it (see Fig.~\ref{fig:singlecloud_morphology}c).  Because
field lines cannot slip around the surface of the cloud (again due to its
nature as an infinite cylinder), magnetic pressure and field tension
continue to build with the result that the cloud accelerates rapidly
downstream (compare the positions of the clouds in
Fig.~\ref{fig:singlecloud_morphology}).  This rapid acceleration acts
to reduce the magnetic pressure and tension. Again we expect the
  evolution to be quite different to that of a spherical cloud.

\subsection{Two-cloud interactions}
\label{sec:2cloud}
We now investigate the interaction of magnetized shocks with 2 closely
positioned clouds. We first examine the morphology of the interaction,
and then discuss  the acceleration of the clouds  and the evolution of
the plasma  $\beta$. The 2-cloud  arrangements are specified  by their
``width'', which is the lateral  distance between the cloud centers in
units of the cloud radius (\ie  \, the separation of the clouds in the
``y'' direction),  and by their ``offset'', which  is the longitudinal
distance  between  the clouds  (i.e.  their  separation  in the  ``x''
direction). $t=0$  is defined as the  time that the  shock reaches the
leading edge of the more upstream of the two clouds.

\begin{figure}
  \begin{center}
      \includegraphics[width=7cm]{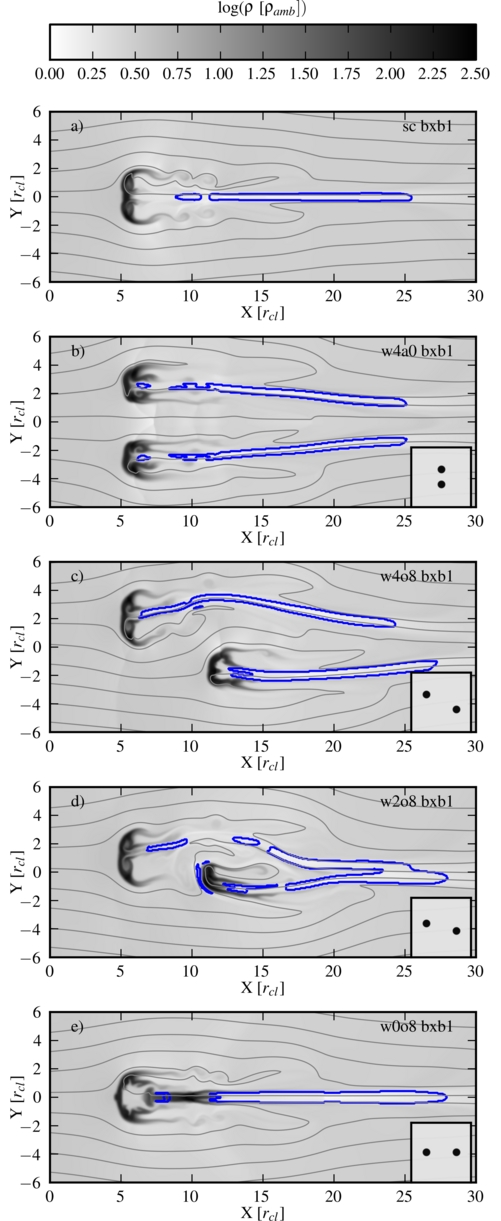} \\
      \caption{Snapshots of  the morphology of a)  an individual cloud
        and  b)-e)  2-clouds with  varying  separation  and offset  at
        $t=4\,t_{\rm cc}$. In all cases the magnetic field is parallel
        to the shock normal and $\beta_0=1.13$. The contour again
        shows the  ``flux rope''  ($\beta<1$ and $\rho  |\mathbf{u}| <
        0.5 \times \rho_{ps}  |\mathbf{u}_{ps}|$), while the grayscale
        shows  the logarithmic  density. The  2-cloud  simulations are
        identified  by the  initial  ``width'' and  ``offset'' of  the
        clouds  - the  relative positions  of the  cloud at  $t=0$ are
        shown in the inset of each panel (shown at reduced scale). The
        resolution is  $R_{32}$. At  higher resolution the  fine scale
        structure changes  somewhat, but  the general features  of the
        flow and  their dependence on  the initial arrangement  of the
        clouds remain unchanged.}
    \label{fig:mor_w4}
    \end{center}
  \end{figure}

\begin{figure}
  \begin{center}
      \includegraphics[width=7cm]{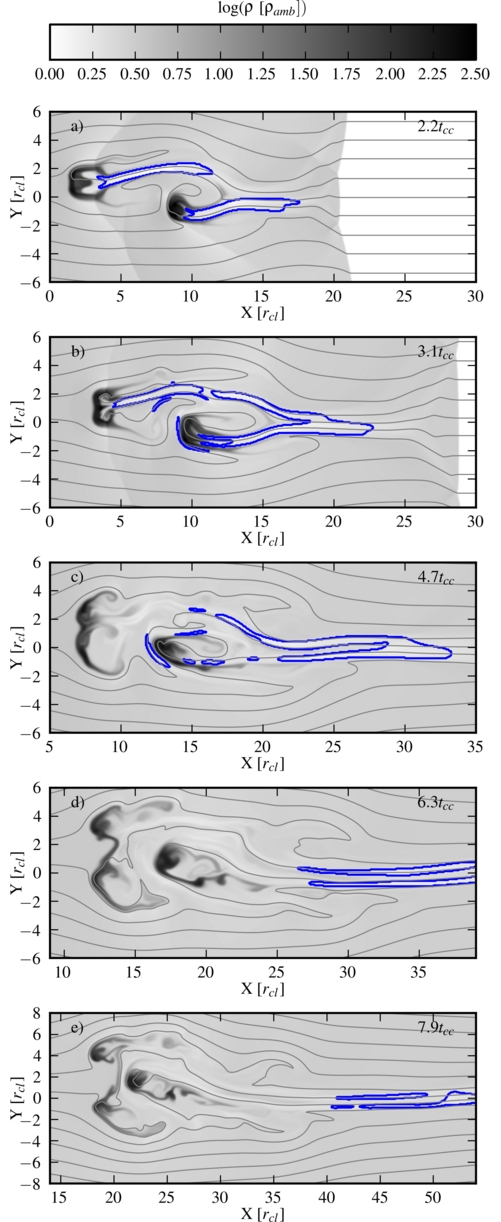} \\
      \caption{The   time   evolution   of  the   2-cloud   simulation
        \emph{s2w2o8}  (the  clouds  are  positioned with  an  initial
        ``width'' $=2  \, r_{cl}$ and ``offset'' $=8  \, r_{cl}$). The
        magnetic   field  is   parallel  to   the  shock   normal  and
        $\beta_0=1.13$.  The logarithmic  density  and magnetic  field
        evolution are  shown at times  $t = 2.2$, $3.1$,  $4.7$, $6.3$
        and $7.9  \, t_{cc}$ (top  to bottom). The contour shows
        the ``flux-rope'' ($\beta<1$ and $\rho |\mathbf{u}|<0.5 \times
        \rho  |\mathbf{u}|_{ps}$). In  this simulation  the downstream
        cloud is confined by the  presence of the upstream cloud. Note
        the changes in the $x$- and $y$-coordinates in each panel.}
    \label{fig:mor_evo}
    \end{center}
  \end{figure}

\subsubsection{Parallel shocks}
\label{sec:2cloud_parallel_morphology}
In interactions with a parallel shock, the presence of a second cloud
alongside the first cloud has the effect of suppressing the lateral
re-expansion of the cloud. This is easily seen when comparing the
single-cloud simulation \emph{sc} and the 2-cloud simulation
\emph{w4o0} (in panels a) and b) of Fig.~\ref{fig:mor_w4},
respectively). The flow between the clouds is slowed and squeezed, but
accelerates once past the clouds.  The initial high pressure between
the clouds drops due to the Bernoulli effect, causing the initial
outwardly directed orientation of the flux-ropes to change towards an
inwardly directed orientation\footnote{This behaviour is also seen in
purely hydrodynamic simulations \citep{2005MNRAS.361.1077P}.}.

As the initial  position of one of the clouds  is moved downstream the
lateral suppression  of the upstream  cloud is reduced and  it evolves
more  like the  single cloud  case. However,  the downstream  cloud is
still much more  affected by lateral confinement (see  the results for
\emph{w4o8} shown in Fig.~\ref{fig:mor_w4}c).

The morphology of  the downstream cloud is dependant  on the ``width''
as  well as  the  ``offset'',  though the  ``width''  is the  dominant
parameter.  The simulations  \emph{w4o8}, \emph{w2o8}  and \emph{w0o8}
shown in panels c)-e) of  Fig.~\ref{fig:mor_w4} illustrate the
diversity of  the downstream  cloud morphology, which  we find  can be
categorised  into three  main types.  When there  is a  sufficient gap
between  the  clouds  for the  flow  to  weave  through (e.g.,  as  in
simulation    \emph{w4o8}    -     see   Fig.~\ref{fig:mor_w4}c), the downstream cloud  is confined in a similar
manner as if there was a cloud alongside it. In contrast, when a cloud
is  directly  behind  an   upstream  cloud  (e.g.,  as  in  simulation
\emph{w0o8} - see Fig.~\ref{fig:mor_w4}e), it falls
in its ``flux rope''. The cloud is shielded from the flow and does not
accelerate. The flow that tries  to converge behind the upstream cloud
(which  forms  the  ``flux   rope'')  instead  now  converges  on  the
downstream cloud, compressing it into an elongated shape. The upstream
cloud is also affected by the  presence of the downstream cloud. As it
accelerates towards the downstream  cloud the tenuous gas between them
is  compressed, modifying  the  morphology of  the  upstream cloud  in
advance of their collision.

The  third type  of  behaviour  occurs when  the  downstream cloud  is
positioned such that  it lies in the ``wings'' of  the flow around the
upstream cloud (e.g.,  see simulation \emph{w2o8} - shown  in panel d)
of  Fig.~\ref{fig:mor_w4}). To  better understand  the nature  of this
interaction  we also  show the  time evolution  of this  simulation in
Fig.~\ref{fig:mor_evo}.  We find that  the ``flux  ropes'' of  the two
clouds  merge downstream,  while the  magnetic field  near  the clouds
becomes highly irregular. The latter  is affected by the fact that the
background flow becomes  quite turbulent as it tries  to force its way
between the  clouds at the same  time as the clouds  are distorted and
influenced by the flow. The turbulent nature of the flow appears to be
quite  efficient  at  stripping  material  away  from  the  downstream
cloud. In spite of this, the  cloud is mostly confined into an $r_{\rm
cl}$-sized   clump  and   does   not  spread   very   far  along   its
fieldlines. Similar behaviour for the downstream cloud is also seen in
simulation \emph{w4o8} at late times as the upstream cloud expands and
the downstream cloud is pushed into the shielded region.

\begin{figure}
  \begin{center}
      \includegraphics[width=7cm]{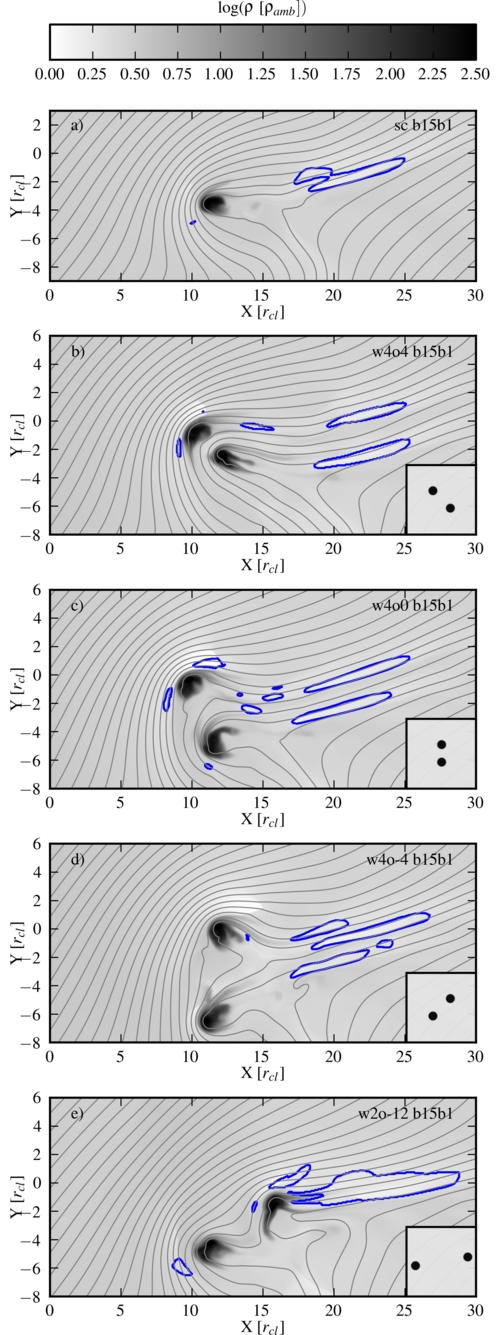} \\
      \caption{Snapshots at $t= 4\,t_{cc}$ of the morphology and field
        structure of shock-cloud simulations with an oblique magnetic
        field ($\theta_0=15^\circ$ and $\beta_0 = 1.13$). The top
        panel shows the interaction with a single cylindrical cloud
        (\emph{sc b15b1}), while the remaining panels show the interaction
        with two cylindrical clouds. The grayscale shows the
        logarithmic density while the contour shows the ``flux
        rope''.}
    \label{fig:mor_w4_ob}
    \end{center}
  \end{figure}

\begin{figure}
  \begin{center}
    \includegraphics[width=7cm]{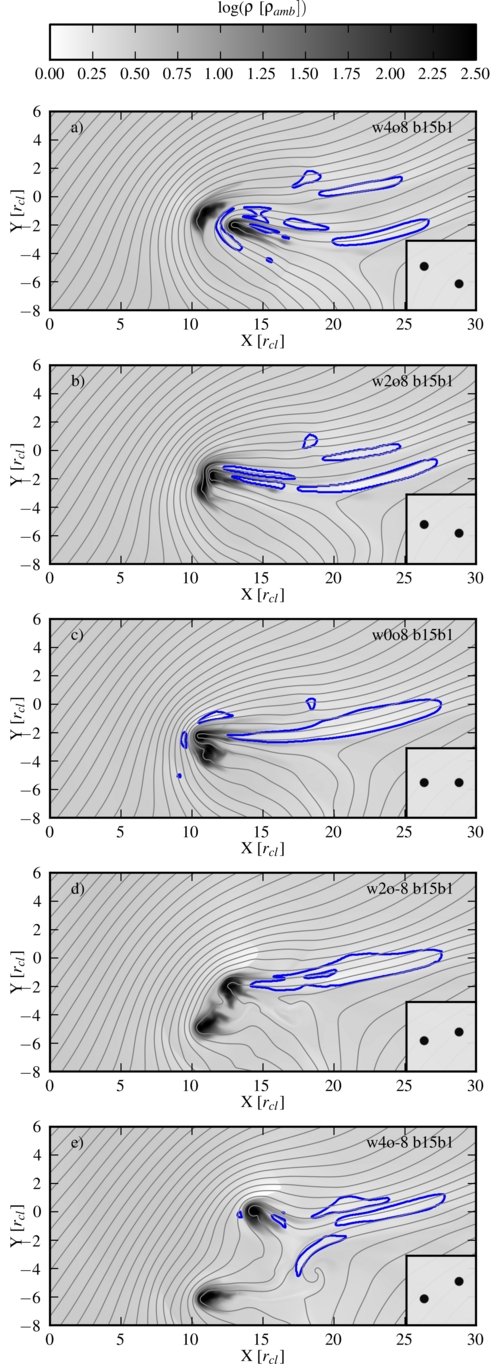} \\  
    \caption{2-cloud oblique-field snapshots like those in
      Fig. \ref{fig:mor_w4_ob} but for a fixed cloud ``offset'' of $8 \, r_{cl}$ and varied ``width''.}
    \label{fig:mor_o8_ob}
    \end{center}
  \end{figure}

\begin{figure}
  \begin{center}
    \includegraphics[width=7cm]{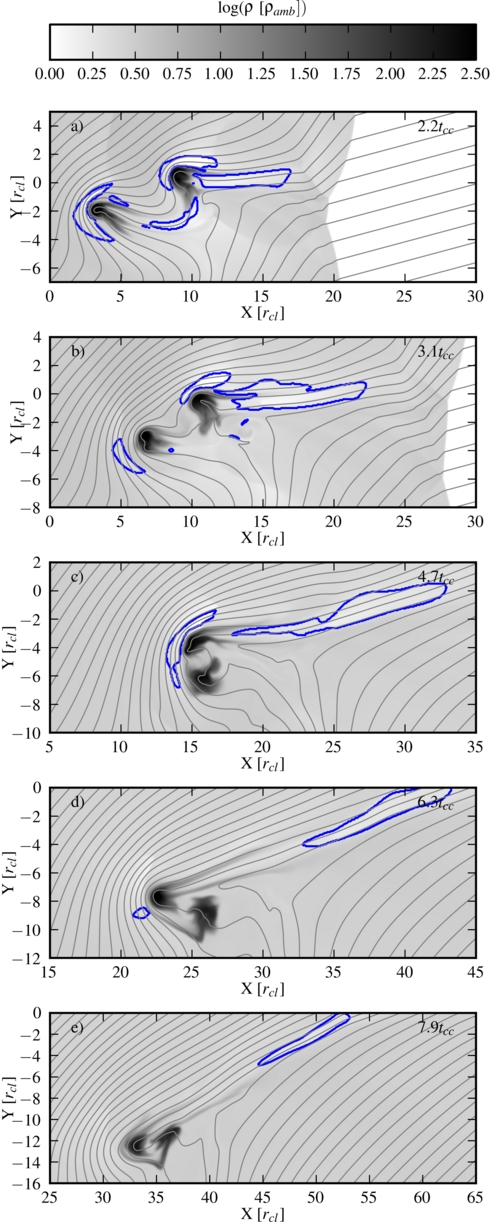} \\  
    \caption{The evolution of the 2-cloud simulation
        \emph{s2w2o-8} (the clouds are positioned with an initial
        ``width'' $=2 \, r_{cl}$ and ``offset'' $=-8 \, r_{cl}$). The
        magnetic field is oblique to the shock normal ($\theta =
        15^\circ$ and
        $\beta_0=1.13$). The logarithmic density and magnetic field
        evolution are shown at times $t = 2.2$, $3.1$, $4.7$, $6.3$
        and $7.9 \, t_{cc}$ (top to bottom). The contour shows
        the ``flux-rope'' ($\beta<1$ and $\rho |\mathbf{u}|<0.5 \times
        \rho |\mathbf{u}|_{ps}$). In this simulation the cloud which
        is initially upstream (i.e. the bottom cloud) is accelerated
        past the top cloud such that it becomes the most downstream
        cloud for $t \gtsimm 4.7 \, t_{\rm cc}$.}
    \label{fig:mor_evo_ob}
    \end{center}
  \end{figure}

\subsubsection{Oblique shocks}
\label{sec:2cloud_oblique_morphology}
We now  study the interaction of  an oblique shock  with 2 cylindrical
clouds.  As the  oblique magnetic  field  is not  symmetric about  the
x-axis it  provides another  direction to supplement  the ``upstream''
and the  ``downstream'' designations. We define  the ``upfield'' cloud
as the  one whose fieldlines encounter  the shock front  first. In the
cases considered the upfield cloud  is almost always the ``top'' cloud
(i.e.  has an  initial positive  ``y'' position).  The  exceptions are
simulations \emph{w2o-8} where the two  clouds lie on roughly the same
fieldlines, and  \emph{w2o-12} which  was chosen specifically  to have
the ``bottom'' cloud as the ``upfield'' one.

Figs.~\ref{fig:mor_w4_ob} and~\ref{fig:mor_o8_ob} compare snapshots of
the density and magnetic field structure of a single cloud case and a
range of two cloud arranagements at $t=4\,t_{cc}$. \emph{Note that a
  negative ``offset'' signifies that the ``top'' cloud is the
  downstream one}.  In all cases the field geometry causes the clouds
to accelerate downwards (to more negative $y$ positions) at the same
time that they are accelerated downstream (to more positive
$x$-positions). We see that the nature of the interaction is
significantly modified by the presence of the second cloud, and that
it depends on the relative initial positions of the clouds. In some
cases the downstream cloud is protected from the oncoming flow by its
position in the lee of the upstream cloud (e.g. as seen in simulation
\emph{w4o4} in Fig.~\ref{fig:mor_w4_ob}, and in simulations
\emph{w4o8} and \emph{w2o8} in Fig.~\ref{fig:mor_o8_ob}). In other
cases the downstream cloud feels the full fury of the oncoming flow
(e.g., as seen for the top cloud in simulation \emph{w4o-4} in
Fig.~\ref{fig:mor_w4_ob} and simulation \emph{w4o-8} in
Fig.~\ref{fig:mor_o8_ob}). Whether the top or bottom cloud accelerates
fastest downstream depends on their relative orientation to the shock
and the field (e.g., in simulation \emph{w4o4} in
Fig.~\ref{fig:mor_w4_ob} and in simulations \emph{w4o8} and
\emph{w2o8} in Fig.~\ref{fig:mor_o8_ob} the top cloud accelerates
fastest downstream, while in simulations \emph{w4o-4} and
\emph{w2o-12} in Fig.~\ref{fig:mor_w4_ob} and simulations \emph{w2o-8}
and \emph{w4o-8} in Fig.~\ref{fig:mor_o8_ob} the bottom cloud does
so).  Note that the bottom cloud in simulation \emph{w0o8} shown in
Fig.~\ref{fig:mor_o8_ob} is initially the upstream cloud.

Because the field lines are now forced to bend around two clouds, in
many cases the region where the magnetic field is parallel to the
direction of the shock propagation becomes larger and another region
where the field is perpendicular extends between the two clouds (see,
e.g., simulations \emph{w4o4}, \emph{w4o0} and \emph{w4o-4} in
Fig.~\ref{fig:mor_w4_ob}). The clouds are also a lot less circular
than compared to the case of a single cloud with an oblique field
(compare any panel in Figs.~\ref{fig:mor_w4_ob}
and~\ref{fig:mor_o8_ob} with panel a) in
Fig.~\ref{fig:mor_w4_ob}).  Stripping now frequently occurs along
multiple directions.

In many cases the wrapping of the field lines cause the top cloud to
accelerated downwards (i.e. to more negative $y$ positions) faster
than the bottom cloud is accelerated in this direction. This can cause
the clouds to either collide or come as close together as allowed by
the magnetic pressure which builds between them (see simulations
\emph{w4o8} and \emph{w2o8} in Fig.~\ref{fig:mor_o8_ob}). In other
cases we find that the upstream cloud can become the most downstream
cloud as the interaction evolves.  Fig.~\ref{fig:mor_evo_ob} shows a
time sequence from simulation \emph{w2o-8b15b1} which shows how the
upstream cloud (in this case the bottom cloud) overtakes the
downstream (top) cloud. Once the bottom cloud moves into the ``lee''
of the top cloud it experiences reduced confinement forces and begins
to diffuse. Simultaneously the top cloud becomes more exposed to the
oncoming flow and experiences another episode of compression. This
type of behaviour is seen in a large range of oblique simulations.

\begin{figure}
  \begin{center}
    \includegraphics[width=7cm]{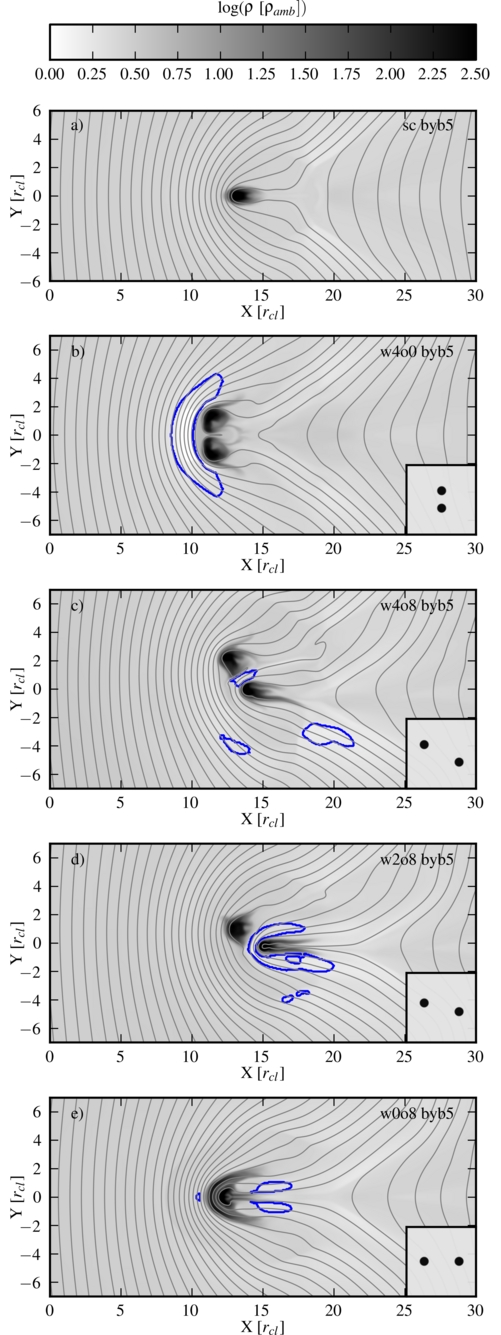} \\  
    \caption{As Fig.~\ref{fig:mor_w4_ob} but with
    perpendicular magnetic fields and $\beta_0 = 5.06$. The time of
    each snapshot is again $t = 4\,t_{\rm cc}$.}
    \label{fig:mor_w4_perp}
    \end{center}
  \end{figure}

\begin{figure}
  \begin{center}
    \includegraphics[width=7cm]{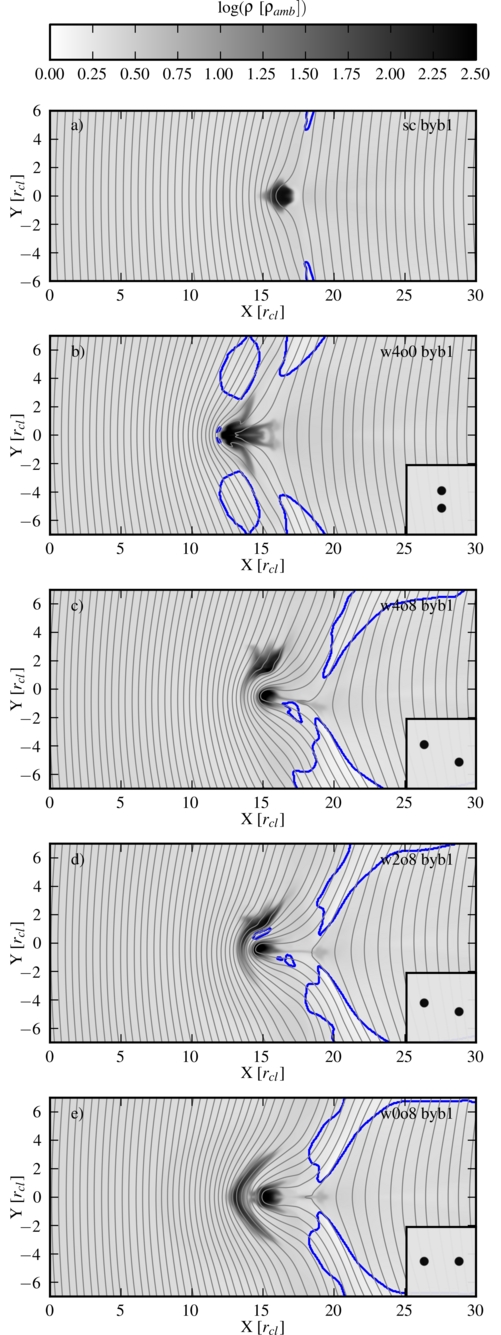} \\  
    \caption{As Fig.~\ref{fig:mor_w4_perp} but with $\beta_0 = 1.13$. The time of
    each snapshot is again $t = 4\,t_{\rm cc}$. The stronger magnetic
    field now controls the dynamics more compared to the simulations
    shown in Fig.~\ref{fig:mor_w4_perp}.}
    \label{fig:mor_w4_perp2}
    \end{center}
  \end{figure}

\begin{figure}
  \begin{center}
    \includegraphics[width=7cm]{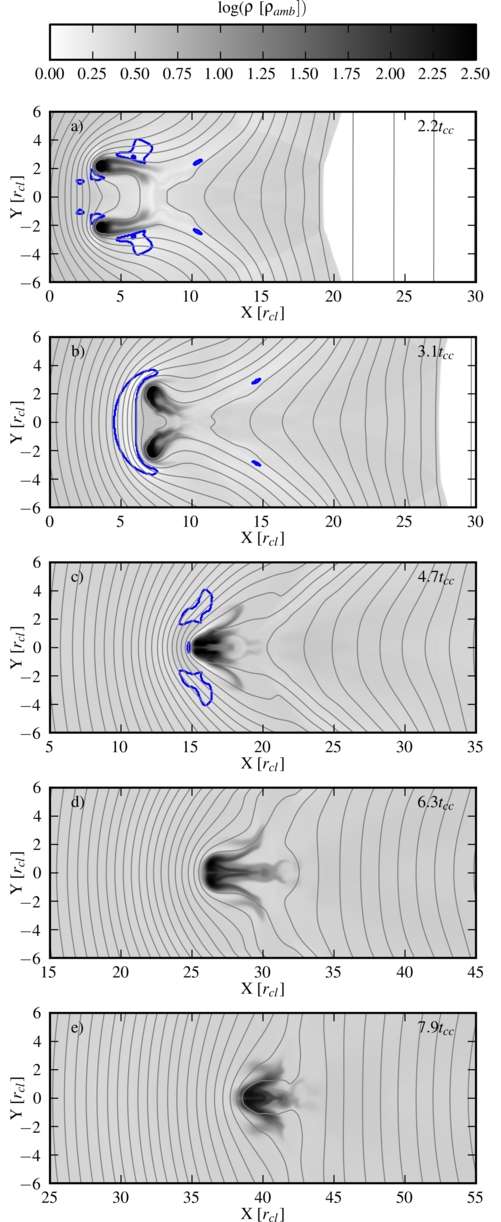} \\  
      \caption{The time evolution of the 2-cloud simulation
        \emph{w4o0\_byb5} (the clouds are positioned with an initial
        ``width'' $=4 \, r_{\rm cl}$ and ``offset'' $=0 \, r_{\rm
          cl}$) The magnetic field is perpendicular to the shock
        normal ($\beta_0=5.06$). The logarithmic density and magnetic
        field evolution are shown at times $2.2$, $3.1$, $4.7$, $6.3$
        and $7.9 \, t_{\rm cc}$ (top to bottom). The contour
        shows the ``flux-rope'' ($\beta<1$ and $\rho |\mathbf{u}|<0.5
        \times \rho |\mathbf{u}|_{ps}$). See also the second panel in Fig.~\ref{fig:mor_w4_perp}.}
    \label{fig:mor_evo1_perp}
    \end{center}
  \end{figure}

\begin{figure}
  \begin{center}
    \includegraphics[width=7cm]{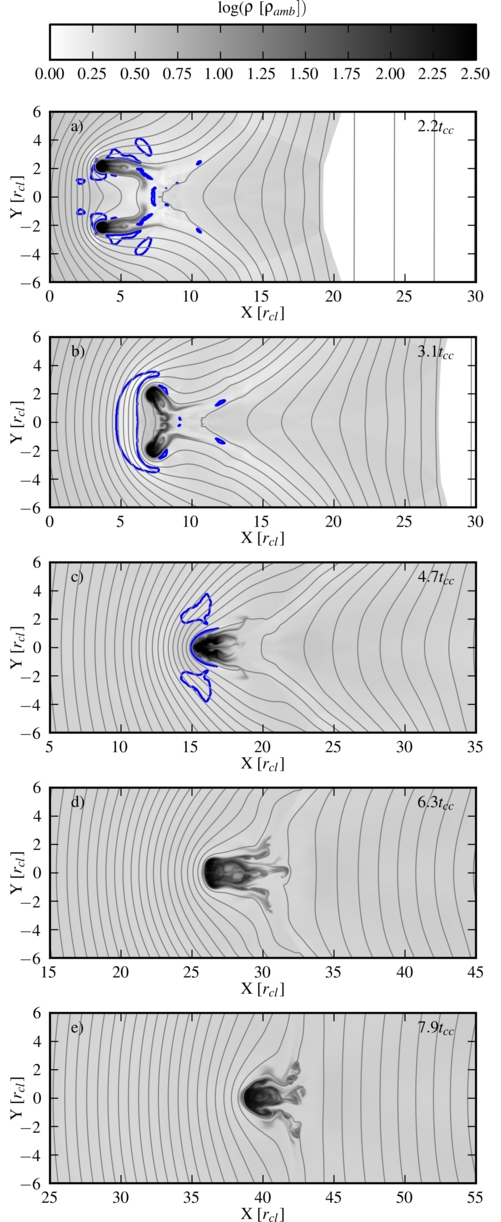} \\  
      \caption{As Fig.~\ref{fig:mor_evo1_perp} but with a resolution
        of 128 cells per cloud radius (instead of 32).}
    \label{fig:mor_evo2_perp}
    \end{center}
  \end{figure}

\begin{figure}
  \begin{center}
    \includegraphics[width=7cm]{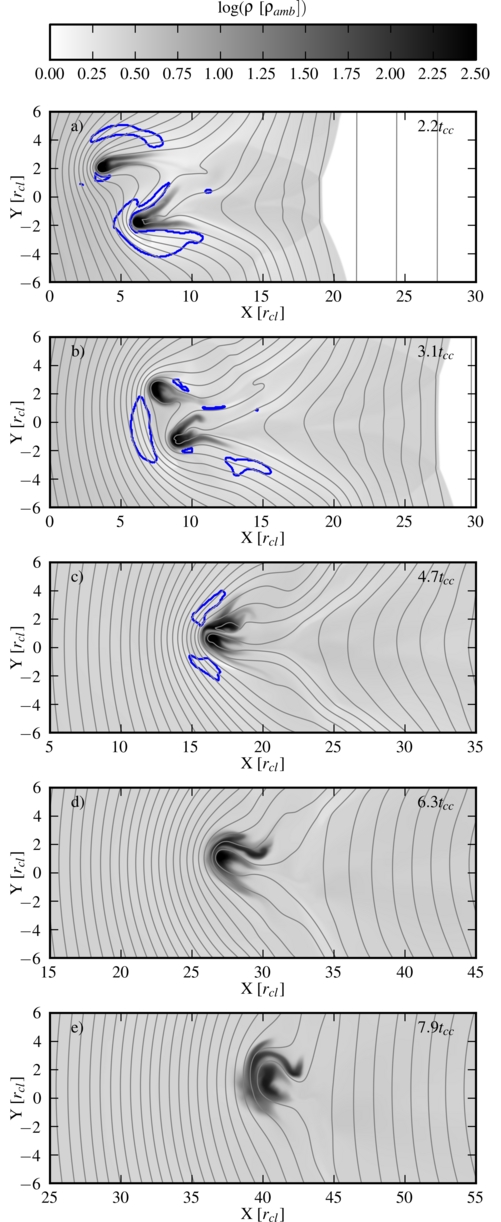} \\  
      \caption{The time evolution of the 2-cloud simulation
        \emph{w4o4\_byb5} (the clouds are positioned with an initial
        ``width'' $=4 \, r_{\rm cl}$ and ``offset'' $=4 \, r_{\rm
          cl}$) The magnetic field is perpendicular to the shock
        normal ($\beta_0=5.06$). The logarithmic density and magnetic
        field evolution are shown at times $t = 2.2$, $3.1$, $4.7$, $6.3$
        and $7.9 \, t_{\rm cc}$ (top to bottom). The contour
        shows the ``flux-rope'' ($\beta<1$ and $\rho |\mathbf{u}|<0.5
        \times \rho |\mathbf{u}|_{ps}$). In this simulation the clouds
        accelerate towards each other with the upstream cloud
        eventually wrapping around the downstream cloud.}
    \label{fig:mor_evo_perp}
    \end{center}
  \end{figure}

\subsubsection{Perpendicular shocks}
\label{sec:2cloud_perpendicular_morphology}
Finally, we study the interaction of a perpendicular shock with two
cylindrical clouds. Figs.~\ref{fig:mor_w4_perp}
and~\ref{fig:mor_w4_perp2} compare snapshots of the density and
magnetic field structure of interactions of a single cloud and 2
clouds with a perpendicular shock at $t=4\,t_{\rm cc}$. In
Fig.~\ref{fig:mor_w4_perp} the plasma $\beta$ of the pre-shock medium
is $\beta_{0} = 5.06$, whereas the field is significantly stronger in
Fig.~\ref{fig:mor_w4_perp2} ($\beta_{0} = 1.13$). As the field
strength increases the magnetic field increasingly controls the
dynamics of the interaction. This is evident from the suppressed
instabilities and cloud mixing, enhanced diffusion of the cloud along
the field lines, greater accceleration of the clouds downstream, and
straighter field lines in Fig.~\ref{fig:mor_w4_perp2} versus
Fig.~\ref{fig:mor_w4_perp}.

We again find that the presence of a second cloud has a major
influence on the nature of the interaction. As the field lines wrap
around the two clouds they are driven towards each other very
rapidly. If clouds lie on the same field line they merge into a single
clump (see the time evolution of simulations \emph{w4o0} in
Figs.~\ref{fig:mor_evo1_perp} and~\ref{fig:mor_evo2_perp}).  During
this process a large continuous region of high magnetic pressure forms
upstream of the clouds. Comparison of Figs.~\ref{fig:mor_evo1_perp}
and~\ref{fig:mor_evo2_perp} reveals that there is some numerical
diffusion present in the $R_{32}$ simulations but that the same
general behaviour occurs\footnote{Due to this difference in numerical
diffusion we find that the degree to which clouds merge when they do
not lie on the same field lines is dependent on the resolution, with
higher resolution simulations better able to prevent mixing and
maintain distinct clouds in such cases (stronger fields also tend to
keep clouds separate). $R_{128}$ resolution is also necessary for
accurate calculation of the plasma $\beta$ in some circumstances - see
Sec.~\ref{sec:2cloud_beta}.}.  If the clouds do not lie on the same
field line then a build up in the magnetic pressure between the clouds
prevents their merger (see simulation \emph{w4o8} in
Fig.~\ref{fig:mor_w4_perp} where the contour between the clouds
highlights the region of high magnetic
pressure). \cite{Lazarian2013} argues that the actual
reconnection diffusion in turbulent plasmas might be quite fast and
there might be a resemblance between numerical diffusion and magnetic
reconnection in turbulent flows.

If the clouds are aligned or nearly-aligned with the direction of
shock propagation the downstream cloud is shielded from the oncoming
flow by the upstream cloud which moves very close towards it (see
simulations \emph{w2o8} and \emph{w0o8} in
Fig.~\ref{fig:mor_w4_perp}).  In such cases, the magnetic field lines
between the clouds prevent the clouds from merging. The downstream
cloud is compressed laterally by the upstream cloud which wraps around
it.

In some cases, clouds which are initially separated quite widely can
be driven towards each other to end up in a very compact
arrangement. This behaviour is shown in Fig.~\ref{fig:mor_evo_perp},
which shows the evolution of the interaction in simulation
\emph{w4o4}. In such cases, shock compression of the field lines
naturally reduces the ``offset'' between the clouds, while their
``width'' is easily reduced by their motion along the field lines. In
this example the downstream cloud moves towards the low pressure
region behind the upstream cloud and away from the high (magnetic)
pressure region around the outside edge of the combined clouds. The
field lines between the clouds prevent complete merging in this
instance.

\begin{figure}
  \begin{center}
    \begin{tabular}{c}
      \includegraphics[width=7cm]{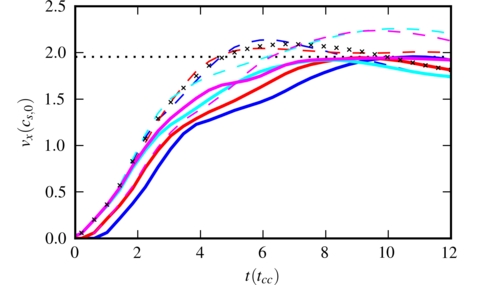} \\
      \includegraphics[width=7cm]{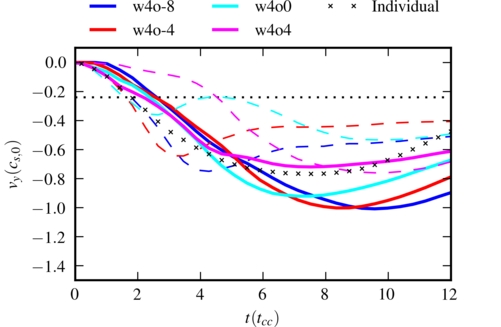} \\
      \end{tabular}
      \caption{Evolution of the \emph{x}- (top panel) and \emph{y}-
        (bottom panel) cloud velocity components in simulations with 2
        clouds and oblique magnetic fields. The velocity is normalized
        by the sound speed of the intercloud ambient medium. The
        initial ``width'' of the cloud distribution is identical in
        each simulation (being $4\, r_{cl}$), while the ``offset'' is
        varied. In each panel the ``top'' cloud in the distribution is
        shown using solid lines while dashed lines correspond to the
        ``bottom'' cloud. The dotted black line shows the intercloud
        velocity of the post-shock flow. Also shown is the velocity
        evolution of a single cloud simulation (indicated by the black
        crosses).}
    \label{fig:v2}
    \end{center}
  \end{figure}

\begin{figure}
  \begin{center}
    \begin{tabular}{c}
      \includegraphics[width=7cm]{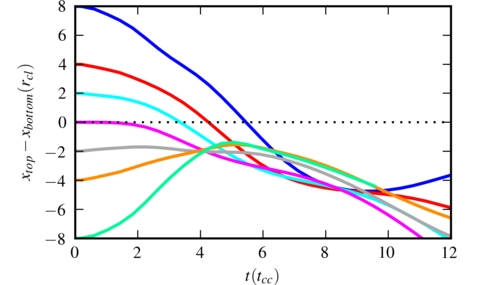} \\
      \includegraphics[width=7cm]{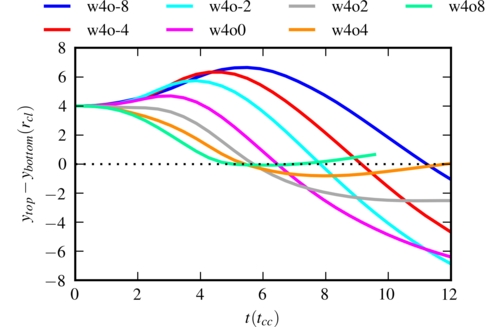} \\

      \end{tabular}
    \caption{The evolution of the \emph{x}- and \emph{y}- separations
      of the clouds in
    2-cloud simulations with oblique magnetic fields. A sign change
    (i.e. movement across the horizontal black line) represents
    a switch in relative position.}
    \label{fig:v2_dr}
    \end{center}
  \end{figure}

\subsubsection{Cloud velocities}
\label{sec:2cloud_vel}
In simulations with a parallel or perpendicular magnetic field the
clouds generally develop a small $y$-component to their velocity
which often draws the clouds towards each other (see, e.g.,
simulation \emph{w2o8} in Fig.~\ref{fig:mor_evo} and simulation
\emph{w4o4\_byb5} in Fig.~\ref{fig:mor_evo_perp}). 

However, the velocity evolution of a cloud is generally far more
significant when the magnetic field is oblique.  A clear and
systematic distinction between the $x$-velocity component of the
``top'' and ``bottom'' clouds can be seen in Fig.~\ref{fig:v2}. The
``upstream'' cloud accelerates first which is the ``top'' cloud for
positive ``offset'' and the ``bottom'' cloud if the ``offset'' is
negative. Initially, the \emph{x}-velocity in the ``bottom'' cloud
grows at a rate similar to the isolated cloud case (compare the dotted
lines for simulations \emph{w4o-8}, \emph{w4o4} and \emph{w4o0} with
the black crosses). The $v_{x}$ velocity of each of these clouds
overshoots slightly the post-shock flow value, as does the isolated
cloud. In contrast, the acceleration of the ``top'' cloud is noteably
slower after about $2.5\,t_{\rm cc}$ and in all simulations it reaches
the post-shock flow value without any overshoot.

The bottom panel of Fig.~\ref{fig:v2} shows the evolution of the
$y$-velocity component of the clouds. In the single cloud case the
cloud significantly overshoots the velocity of the postshock flow
which has a normalized value $v_{y} \approx -0.25\,c_{\rm s,0}$. The
single cloud reaches its peak $y$-velocity of $\approx -0.8\,c_{\rm
  s,0}$ at $t \approx 7.5\,t_{\rm cc}$, before decelerating. At late
times we would expect the cloud $v_{y}$ to asymptote towards that of
the postshock flow but this clearly takes place on timescales in
excess of $12\,t_{\rm cc}$. The $y$-velocity component of the clouds
in the 2-cloud simulations follows the same broad behaviour of initial
acceleration, overshoot of the equilibrium value, and deceleration
towards the postshock speed, but there are significant differences in
the details. The ``top'' cloud accelerates downward slowly initially,
but significantly overshoots the isolated cloud case later on (unless
the ``top'' cloud is also the ``upstream'' one (e.g., \emph{w4o4}), in
which case its behaviour is closer to the isolated cloud). In contrast
the ``bottom'' cloud initially accelerates faster than the isolated
cloud, but starts slowing down much sooner (reaching a peak velocity
of $\approx -0.65\,c_{\rm s,0}$ at $t \approx 3\,t_{\rm cc}$ for
\emph{w4o-4}). Simulation \emph{w4o4} is again the exception - as the
``downstream'', ``bottom'' cloud is shielded from the flow it
accelerates very slowly initially. Finally we note that some clouds
(e.g., the ``bottom'' cloud in simulation \emph{w4o0}) undergo a
second period of acceleration.

Overall, we find that the ``bottom'' cloud moves faster in the ``x''
direction and the ``top'' cloud moves faster in the ``y''
direction. Thus if initially the ``upstream'' cloud is the ``bottom''
one then the upstream cloud will overtake the downstream cloud. This
is highlighted in the top panel of Fig.~\ref{fig:v2_dr} where we see
that the clouds swap relative positions (i.e. cross the horizontal
black line) in simulations \emph{w4o-8}, \emph{w4o-4}, \emph{w4o-2}
and \emph{w4o-1}.  It is also observed in simulation \emph{w2o-8} as
shown in Fig.~\ref{fig:mor_evo_ob}.

However, we also find that the ``top'' and ``bottom'' clouds swap
their relative $y$-positions in all of the simulations with
``width''~$ = 4\, r_{cl}$ that we have investigated. This is shown in the lower
panel of Fig.~\ref{fig:v2_dr} where all the simulations cross the
horizontal black line, irrespective of the initial ``offset''.  We
observe that a swap-over even occurs in simulations like \emph{w4o-8},
where the ``bottom'' cloud is the first to accelerate and the
separation between the clouds actually grows until $6\,t_{\rm cc}$ (in
this case the swap-over occurs at $t > 10\,t_{\rm cc}$).
Fig.~\ref{fig:mor_evo_ob} shows the swap-over process occuring in
simulation \emph{w2o-8} at $t \approx 8\,t_{\rm cc}$ (here the
``bottom'' cloud moves underneath and then behind the ``top'' cloud).

\begin{figure}
  \begin{center}
    \includegraphics[width=7cm]{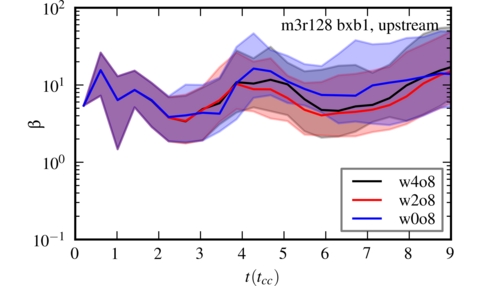}
    \includegraphics[width=7cm]{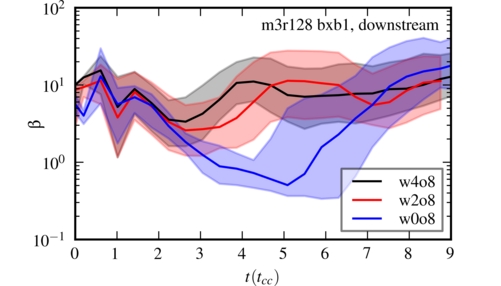}
    \caption{The time evolution of the $\beta$ distributions for
      upstream (top panel) and downstream (bottom panel) clouds in $R_{128}$
      2-cloud simulations with parallel magnetic fields and preshock
      $\beta_0=1.13$. The initial cloud ``offset'' is 8 while the
      initial cloud ``width'' is varied. The solid line shows the
      median $\beta$ value and the area between the 25$^{th}$ and
      75$^{th}$ percentiles is shaded.}
    \label{fig:s2o8}
  \end{center}
\end{figure}

\begin{figure}
  \begin{center}
    \begin{tabular}{c}
      \includegraphics[width=7cm]{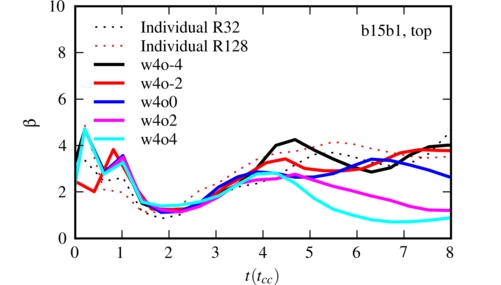} \\
      \includegraphics[width=7cm]{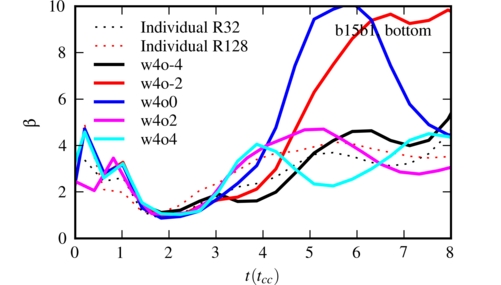} \\

      \end{tabular}
    \caption{Evolution of the harmonic average of $\beta$ in material
      from the ``top'' cloud (top panel) 
      and the ``bottom'' cloud (bottom panel) in 2-cloud simulations
      with an oblique magnetic field (where $\beta_0=1.13$ and
      $\theta_0 = 15^\circ$). The initial cloud positions have a
      ``width'' of 4$\,r_{\rm cl}$ and varying ``offset''. The
      evolution of $\beta$ in isolated clouds is also shown (for
      simulations with 32 ($R_{32}$) and 128 ($R_{128}$) cells per cloud radius).}
    \label{fig:hav_s2w4_obl}
    \end{center}
  \end{figure}

\begin{figure}
  \begin{center}
    \begin{tabular}{c}
      \includegraphics[width=7cm]{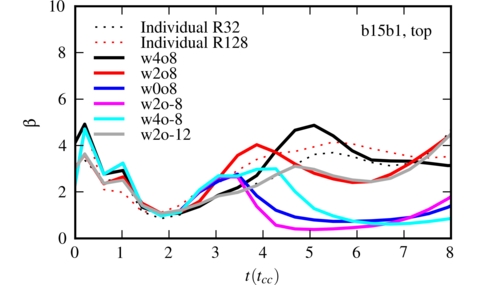} \\
      \includegraphics[width=7cm]{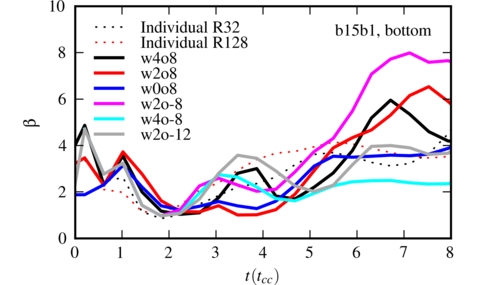} \\
      \end{tabular}
    \caption{As Fig.~\ref{fig:hav_s2w4_obl} but for clouds in
    simulations with an initial ``offset'' of 8$\,r_{\rm cl}$ and
    varying ``width''. The upstream cloud is identified as the ``top''
  cloud in simulation \emph{w0o8}.}
    \label{fig:hav_s2o8_obl}
    \end{center}
  \end{figure}

\begin{figure}
  \begin{center}
    \begin{tabular}{c}
      \includegraphics[width=7cm]{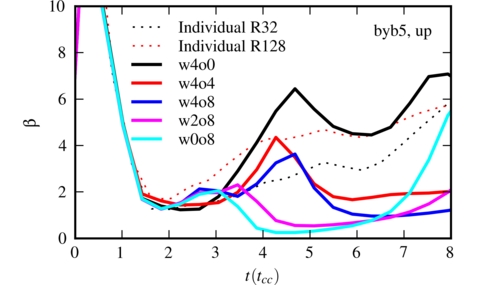} \\
      \includegraphics[width=7cm]{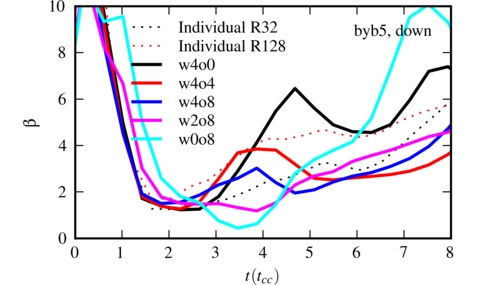} \\

      \end{tabular}
      \caption{Evolution of the harmonic average of $\beta$ in
        material from the upstream (top panel) and downstream (bottom
        panel) cloud in 2-cloud simulations with a perpendicular
        magnetic field ($\beta_0=5.06$). The
      evolution of $\beta$ in isolated clouds is also shown (for
      simulations with 32 ($R_{32}$) and 128 ($R_{128}$) cells per cloud
      radius).}
    \label{fig:hav_s2_perp}
    \end{center}
  \end{figure}

\subsubsection{The plasma $\beta$}
\label{sec:2cloud_beta}
Of the simulations performed, the parallel shock simulations with
$\beta_{0} = 1.13$ (i.e. models \emph{bxb1}) have the highest post-shock $\beta$ ($\sim
\rm 12$, see Table~\ref{table:sims}). It is in these simulations that
instabilities are least suppressed by the magnetic field. Simulations
with single clouds reveal that the results are sensitive to
resolution, with a convergence study indicating that of order 100
cells per cloud radius are needed for accurate results \citep[in keeping
with previous work of adiabatic hydrodynamical shock-cloud
interactions - see, e.g.,][]{1994ApJ...420..213K,2009MNRAS.394.1351P}.
In contrast, the presence of additional clouds disturbs the
flow such that longer wavelength instabilities play a more important
role. This reduces the resolution requirements in multi-cloud
simulations. However, in order to compare like-with-like, we perform
the following analysis of $\beta$ in the parallel shock simulations
using resolution $R_{128}$ for the multi-cloud simulations too.

We first study how the distribution of $\beta$ in the simulations with
a parallel shock changes as the initial positions of the clouds are
varied. In each of the following figures we show the time evolution of
the distribution of the plasma $\beta$ of the cloud material (the
distribution is calculated over all cells in the simulation upstream
of the shock front but is weighted by the amount of cloud material in
each cell). $\beta$ changes with time as the cloud is first
compressed, and then re-expands. At late times $\beta$ should approach
the value in the post-shock flow. This behaviour can be seen in
Fig.~\ref{fig:s2o8}.

We find that varying the initial cloud ``offset'' has no real effect
on the $\beta$ distributions when the initial cloud ``width'' is
greater than the diameter of the clouds. In Fig.~\ref{fig:s2o8} we
show how the evolution of $\beta$ depends instead on the initial
``width'' of the cloud distribution for simulations with
$\beta_0=1.13$. We find that the upstream cloud is not affected in the
\emph{w2o8} simulation, but the growth of $\beta$ is delayed by
$1\,t_{\rm cc}$ in the downstream cloud (compare the red and blue
lines in the bottom panel of Fig.~\ref{fig:s2o8} between $3 \ltsimm
t/t_{\rm cc} \ltsimm 5$). Note, though that this delay is not seen
in the \emph{bxb0.5} case where the magnetic field is more dominant.

In the \emph{w0o8} case (see Fig.~\ref{fig:mor_w4}e)), the downstream cloud falls inside the flux
rope and $\beta$ drops to $\sim 0.5$ in the downstream cloud until the
clouds collide. The beta distribution of the upstream cloud is also
affected in this case - $\beta$ is generally slightly higher due to
the increased pressure downstream. The same behaviour is seen if the
magnetic field is made slightly stronger. For example, in simulations
with $\beta_0=0.55$ (models \emph{bxb0.5}) the minimum $\beta$ is
still around $0.5$ in the downstream cloud, while the increase of the
plasma $\beta$ in the upstream cloud is even more prominent.

We find that simulations with an oblique magnetic field are much less
sensitive to resolution, and we are able to use simulations with a
resolution of 32 cells per cloud radius.  We adopt the
harmonic mean as the average for the $\beta$ statistics in these
simulations: it demonstrates good convergence because it is not
influenced by a small number of cells with high $\beta$ where the flow
is poorly resolved. The harmonic mean is thus a good estimator for the
``typical'' $\beta$ value of cloud material, and it generally falls
inbetween the 30$^{\rm th}$ and 50$^{\rm th}$ percentile values.
 
Figs.~\ref{fig:hav_s2w4_obl} and~\ref{fig:hav_s2o8_obl} show the
evolution of the harmonic mean of $\beta$ in material from the ``top''
and ``bottom'' clouds of various simulations. The ``top'' cloud is the
upstream one if the ``offset'' is positive, and is the ``upfield''
cloud in all simulations except \emph{w2o-12} and \emph{w2o-8}. These
figures also show the variation of $\beta$ in simulations with a
single individual cloud. In Fig.~\ref{fig:hav_s2w4_obl} we see the
effect of varying the ``offset'' value of the initial cloud
distribution while keeping the initial distribution ``width'' fixed at
a value of $4\,r_{\rm cl}$. In contrast, in
Fig.~\ref{fig:hav_s2o8_obl} the initial distribution ``width'' is
varied while the ``offset'' is kept at $8$ or $12r_{cl}$.

These figures reveal that $\beta$ is significantly reduced in the
``top'' cloud when it is the upstream one (see models \emph{w4o2} and
\emph{w4o4} in Fig.~\ref{fig:hav_s2w4_obl}, and models \emph{w4o8},
\emph{w2o8} and \emph{w0o8} in Fig.~\ref{fig:hav_s2o8_obl}).  In model
\emph{w2o8} we see that $\beta < 1$ during the period $4 \ltsimm t/t_{\rm cc}
\ltsimm 7$; Fig.~\ref{fig:mor_o8_ob} shows that the clouds collide at
this time.  In fact, the collision of the clouds is responsible for the
low $\beta$ values in the material of the top cloud in all of these
simulations, and also in simulation \emph{w0o8} (where low $\beta$
values occur in the upstream cloud). In contrast, we find that $\beta$
in material in the ``bottom'' cloud is similar to that in the isolated
cloud or slightly higher.

When the ``top'' cloud is the ``downstream'' one, the harmonic mean of
$\beta$ in both of the clouds evolves similarly to the evolution of
$\beta$ in an isolated cloud. Exceptions to this behaviour occur only
for the bottom cloud in simulations \emph{w4o-2} and \emph{w4o0} (see
Fig.~\ref{fig:hav_s2w4_obl}) and simulation \emph{w2o-8} (see
Fig.~\ref{fig:hav_s2o8_obl}); in these cases the ``bottom'' cloud
reaches much higher $\beta$ values. The reason for this difference is
evident from Fig.~\ref{fig:mor_evo_ob}, which reveals that in
simulation \emph{w2o-8} the ``bottom'' cloud overtakes the ``top''
cloud and becomes the ``downstream'' cloud at the time when $\beta$
starts growing. The same behaviour also occurs in the other two
cases. For example, in simulation \emph{w4o0} the bottom cloud crosses
a line perpendicular to the upstream field lines passing through the
``top'' cloud at this time. Finally, we note that although the clouds
also pass each other in \emph{w4o-4}, this happens at a later time and
greater separation with the result that $\beta$ does not grow as much
in the bottom cloud.

Finally we study the evolution of $\beta$ in simulations with a
perpendicular magnetic field. The $\beta$ in the post-shock flow of
models \emph{byb5} is $6.05$. Since this is the same as
in models \emph{b15b1}, $\beta$ in the shocked clouds varies in the range of
$4-7$ for the majority of cloud arrangements in simulations with these
field values.

The ``upstream'' clouds in simulations \emph{byb5} correspond to
``upstream''-''top'' clouds in the oblique simulations \emph{b15b1}
and thus all such clouds have reduced $\beta$ values (see models
\emph{w4o4}, \emph{w4o8}, \emph{w2o8} and \emph{w0o8} in
Fig.~\ref{fig:hav_s2_perp}). We also find again that $\beta$ in the
downstream clouds evolves similarly to that in isolated clouds, and
that only clouds that are shielded from the flow (such as the
downstream clouds in simulations \emph{w2o8} and \emph{w0o8}) go
through a phase of significantly reduced $\beta$ (occuring at $t
\approx 3-4\,t_{cc}$ in these cases). Because the clouds in simulation
\emph{w4o0} are on the same field line, $\beta$ increases as they
mix. An increase in $\beta$ is also seen in the downstream cloud of
\emph{w0o8} but further examination indicates that it is principally
due to mixing from numerical diffusion as this behaviour is not seen
at higher resolution. Other higher resolution results track the lower
resolution results almost exactly.

\begin{figure}
  \begin{center}
      \includegraphics[width=7cm]{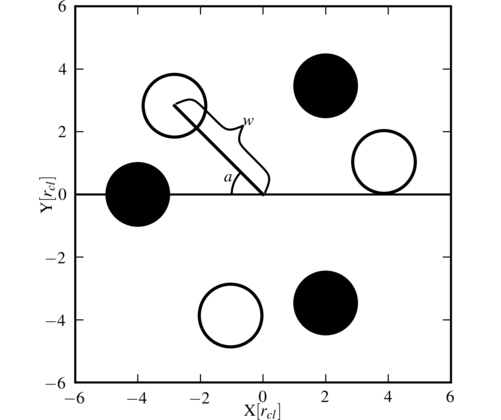} \\
    \caption{Illustrations of the cloud positions in 3-cloud
      simulations. Two particular
    arangements are shown: \emph{s3w4a0} (with the clouds indicated by
    the filled circles) and \emph{s3w4a45} (with the clouds indicated
    by the open circles).}
    \label{fig:scheme}
    \end{center}
  \end{figure}

\begin{figure}
  \begin{center}
    \includegraphics[width=7cm]{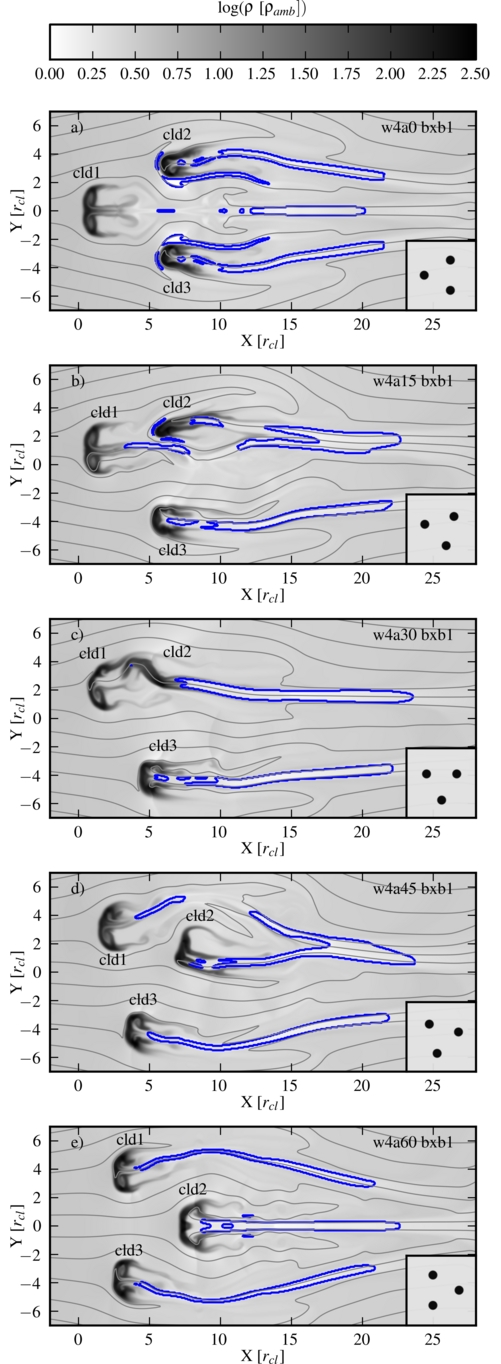} \\
    \caption{Snapshots at $t = 4\,t_{\rm cc}$ of various 3-cloud
    simulations with parallel magnetic fields ($\beta_0 =
    1.13$). Individual clouds are labelled and the insert shows the
    initial cloud arrangement in each case. Only the orientation of
    the cloud arrangement is changed in these cases.}
    \label{fig:mor_s3w4}
    \end{center}
  \end{figure}

\begin{figure}
  \begin{center}
    \includegraphics[width=7cm]{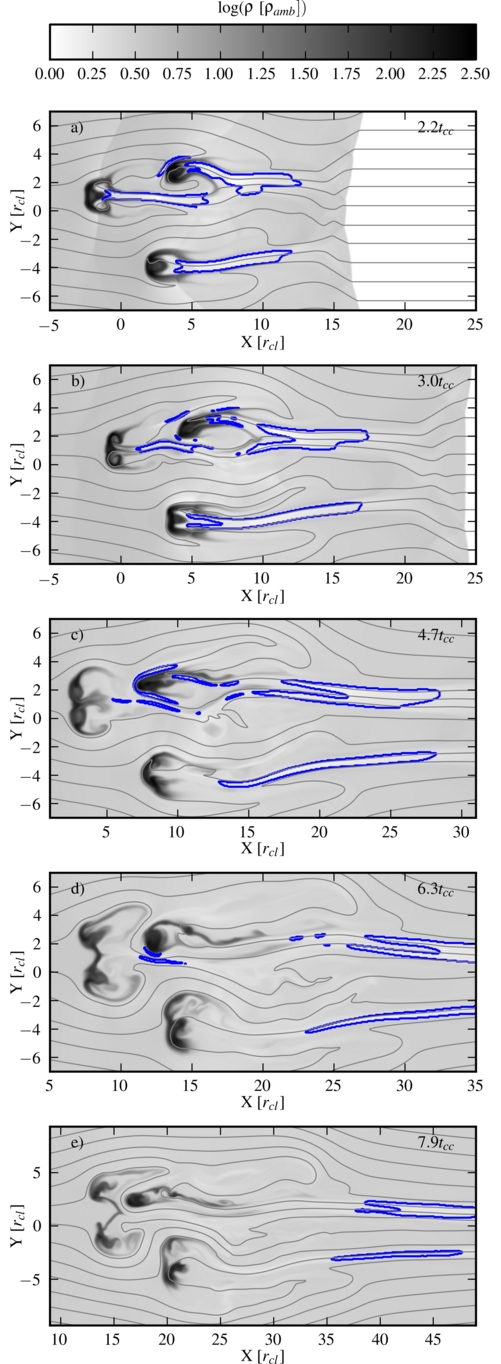} \\
    \caption{The time evolution of the 3-cloud simulation \emph{s3w4a15}
    with a parallel magnetic field ($\beta_0 = 1.13$). The logarithmic density and magnetic
        field evolution are shown at times $t = 2.2$, $3.0$, $4.7$, $6.3$
        and $7.9 \, t_{\rm cc}$ (top to bottom). The contour
        shows the ``flux-rope'' ($\beta<1$ and $\rho |\mathbf{u}|<0.5
        \times \rho |\mathbf{u}|_{ps}$).}
    \label{fig:mor_s3_evo}
    \end{center}
  \end{figure}

%\afterpage{\clearpage} %Try to flush those floats

\begin{figure}
  \begin{center}
    \begin{tabular}{c}
      \includegraphics[width=7cm]{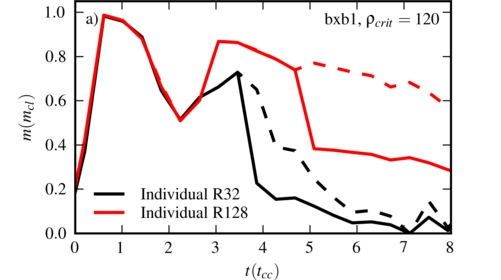} \\
      \includegraphics[width=7cm]{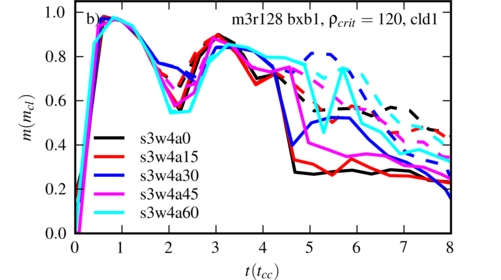} \\
      \includegraphics[width=7cm]{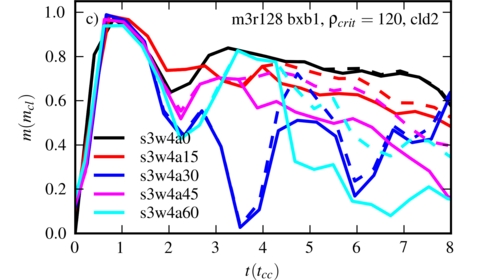} \\
      \includegraphics[width=7cm]{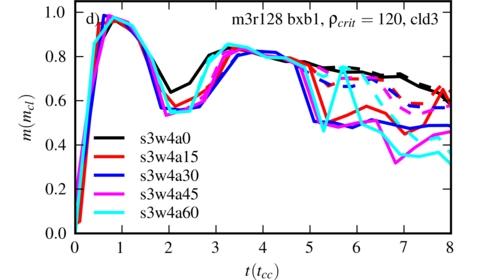} \\
      \end{tabular}
    \caption{Evolution of the core mass (see text) for a) single cloud
    simulations at two different resolutions, and for b) cld1, c) cld2 and d) cld3 in
    high resolution 3-cloud simulations. In each case the solid line represents
    the main fragment and the dashed line shows the sum of all
    fragments. The $t=0$ time for each cloud starts when the shock
    first reaches the cloud.}
    \label{fig:core_mass}
    \end{center}
  \end{figure}

\begin{figure}
  \begin{center}
    \includegraphics[width=7cm]{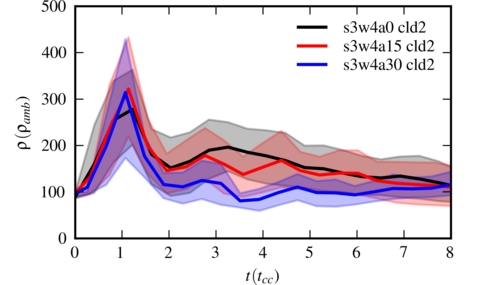} \\
    \caption{Evolution of the density in cld2 in some of the 3-cloud
    simulations. The average density within cld2 is shown by the solid line and the
    region between the 25$^{th}$ and 75$^{th}$ percentiles is shaded.}
    \label{fig:core_density}
    \end{center}
  \end{figure}

\begin{figure}
  \begin{center}
    \begin{tabular}{c}
      \includegraphics[width=7cm]{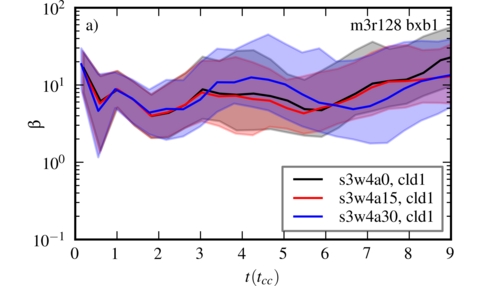} \\
      \includegraphics[width=7cm]{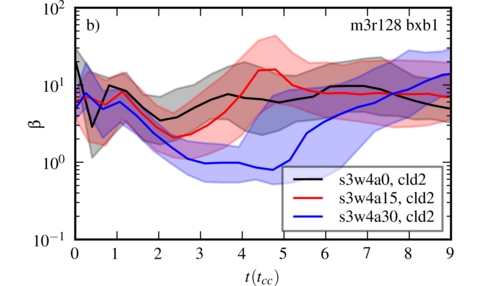} \\
      \includegraphics[width=7cm]{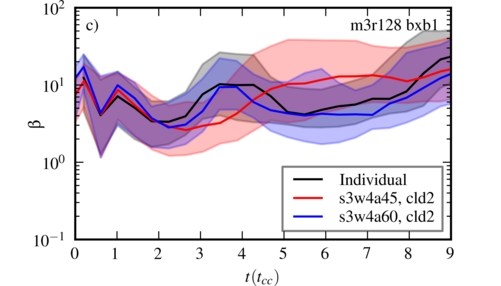} \\
      \includegraphics[width=7cm]{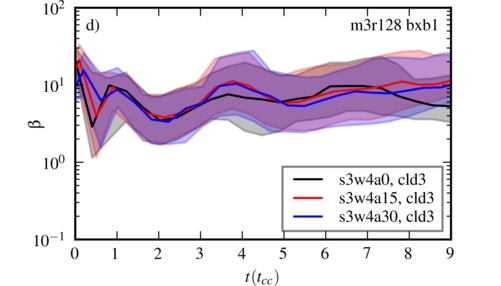} \\
    \end{tabular}
    \caption{The time evolution of the $\beta$ distributions for different
    clouds in high resolution ($R_{128}$) 3-cloud simulations with parallel magnetic
    fields and a preshock $\beta=1.13$. The solid line shows the median
    value and the area between the 25$^{th}$ and 75$^{th}$ percentiles is shaded.}
    \label{fig:s3_r128}
  \end{center}
\end{figure}

\subsection{Three-cloud interactions}
\label{sec:3clouds}
We now investigate the MHD interaction of a shock with 3 closely
spaced clouds which are arranged to form the vertices of an
equilateral triangle (see Fig.~\ref{fig:scheme}). The centroid of the
triangle is located at the origin of the computational grid and the
exact arrangement is defined by the angle between the vector to the
most upstream cloud and the (negative) $x$-axis and the length of this
vector (so distribution \emph{w4a30} has the most upstream cloud
located at $(x,y) = (-4 \times \cos{30^\circ}, 4\times \sin{30^\circ})
= (-3.46, 2)$).  The most upstream cloud is referred to as
``cld1''. The next cloud clockwise, referred to as ``cld2'', will be
the one that is behind (directly or with some lateral offset)
``cld1''.  The final cloud, ``cld3'', is then located off to the side.

A compact, \emph{w4} arrangement gives a side length of
$l=\sqrt{3} \times 4=6.93 \, r_{cl}$ for the equilateral triangle. If
considered as part of a hexagonal lattice this distribution would give
a mass ratio (the ratio of mass in the clouds to the intercloud mass)
$MR=9.07$. A slightly wider \emph{w8} arrangement (not considered in
this work) gives $l=\sqrt{3} \times 8=13.86 \, r_{cl}$ and
$MR=2.12$. The mass ratio can be increased by reducing $w$ and by
increasing the cloud density contrast, $\chi$.

We now investigate the nature of the interaction with parallel,
oblique and perpendicular shocks in turn.

\subsubsection{Parallel shocks}
\label{sec:3clouds_parallel}
The interaction of a shock with 3-clouds can be thought of as being
similar to a 2-cloud scenario, but with the addition of a ``modifier''
cloud. Fig.~\ref{fig:mor_s3w4} shows the nature of the interaction for a
relatively compact arrangement of clouds. When clouds are placed
further apart the morphology of the interaction increasingly resembles
either \emph{w4a0} or \emph{w4a60}, except when the orientation is
such that the clouds line up.

As with the previous 2-cloud simulations, the nature of the 3-cloud
interaction depends on the relative positioning of the clouds.
In Fig.~\ref{fig:mor_s3w4}a), we see that the ``flux
rope'' from cld1 passes inbetween the two downstream clouds and
completely detaches. In addition, an interesting low $\beta$, low
momentum region forms near the inside ``wing'' of the downstream
clouds. Rotating the cloud distribution to break the lateral symmetry
we observe that the ``flux ropes'' of two of the clouds may merge (as
seen in simulations \emph{w4a15} and \emph{w4a45} in
Fig.~\ref{fig:mor_s3w4}b) and d). The merging of flux ropes was previously
seen in the 2-cloud simulation \emph{w2o8} shown in
Fig.~\ref{fig:mor_w4}d). The location of the third cloud influences
the sections of ``flux rope'' associated with individual clouds but
the merged part looks the same.  Finally, when cld2 falls directly
into the ``flux rope'' of cld1 (as seen in simulation \emph{w4a30}
in Fig.~\ref{fig:mor_s3w4}c), the resulting ``flux rope'' appears very
similar to that in the 2-cloud simulation \emph{w0o8} shown in
Fig.~\ref{fig:mor_w4}e), but the morphology of cld2 is
significantly changed by the presence of the third cloud.

The time evolution of simulation \emph{w4a15} is shown in
Fig.~\ref{fig:mor_s3_evo}. In this simulation the strongest
interaction occurs between those clouds with the smallest difference in
their lateral positions (cld1 and cld2 in this case). Compared to
cld2, cld3 is able to retain a broadly symmetric structure for longer,
with the only significant deviations by $t = 3\,t_{\rm cc}$ being to
its tail. After this time, cld3 becomes increasingly asymmetric in
appearance.
At $t = 6\,t_{\rm cc}$, cld2 has a circular core and a tail of stripped
material extending from its outside edge. Such a tail only occurs when
a downstream cloud is in the ``wings'' of an upstream cloud.

To better understand the nature of the interactions between clouds in
the 3-cloud simulations we now look at the evolution of the mass of
the core region of each cloud and each cloud's density.  We define
cloud cores as circular regions with an average density $\langle \rho
\rangle > \rho_{crit} = 120\rho_{amb}$ (i.e. a 20\% increase on the
initial cloud density).  Fig.~\ref{fig:core_mass} shows the evolution
of the core mass in single-cloud simulations and in the 3-cloud
simulations shown in Fig.~\ref{fig:mor_s3w4}. The core mass rises
rapidly as each cloud is compressed and abruptly plateaus once 100\%
of the cloud material is above the density threshold. This takes
roughly one cloud-crushing timescale by definition. Subsequent
re-expansion of each cloud causes the core mass to decrease (in the
single cloud case the core mass decreases to $\approx 0.5\,m_{\rm
  cl}$ by $t\approx 2\,t_{\rm cc}$). In many cases the subsequent
behaviour is oscillatory as the cloud cycles through phases of
expansion and contraction, though a steady decline in the core mass is
the dominant trend as material from the cloud mixes in with the
ambient flow (ultimately the cloud density becomes equal to the
post-shock density).

In many simulations the cloud fragments into multiple cores. When this
happens the mass of the largest fragment is shown by the solid lines in
Fig.~\ref{fig:core_mass} while the sum of the mass of all fragments is
shown by the dotted lines. Any overlapping cores are merged into a single
fragment. We find that this analysis is dependent on the resolution
adopted in the simulations. As shown in the top panel of
Fig.~\ref{fig:core_mass}, a lower resolution simulation diverges from
a higher resolution simulation at $t\approx3\,t_{\rm cc}$.  Therefore
we only consider high resolution runs in this analysis (differences
due to the resolution can be delayed by choosing a lower density
threshold, $\rho_{crit}$). In the high resolution single cloud case, the core splits
into two fragments at $t\approx 5\,t_{\rm cc}$, both of which dip
below $\rho_{crit}$ at $t\approx 6.5\,t_{\rm cc}$ (causing the core mass
shown in Fig.~\ref{fig:core_mass}a) to drop to zero). Subsequent
compression brings material above the density threshold again by
$t\approx 7\,t_{\rm cc}$.

Since cld1 is not downstream of any other cloud, it evolves similarly
to an isolated cloud and fragments at $t\approx 4.5\,t_{cc}$ (see
Fig.~\ref{fig:core_mass}b).  Fragmentation of cld1 is slightly
suppressed in simulation \emph{w4a60} because of the presence of the
other clouds.  alongside. However, subsequent oscillations in the core
mass of cld1 due to expansion and contraction of the cloud appear to
be much weaker compared to the single cloud case, indicating that the
presence of the other clouds is again being felt. At $t = 9 \, t_{\rm
  cc}$, $0.4 \, m_{cl}$ remains in the combined fragments of
cld1. The exception to this is simulation \emph{w4a30}, where the
interaction of cld1 with cld2 pushes the average density of
cld1 down to $70 \, \rho_{amb}$ (i.e. below the density threshold
for identification of material as ``core''). The average density of
cld1 in the other simulations is $\approx 90\, \rho_{amb}$ at this
time, and for simulations with an isolated cloud it is at $\approx 100
\, \rho_{amb}$.

Various types of interaction show up in the behaviour of the core mass
of ``cld2''. Simulations \emph{w4a0} and \emph{w4a15} are noticeable
for the large mass fraction which remains in the core and the lack of
significant fragmentation.  In both these simulations cld2 is on the
``outside'' edge of the distribution, and the average density of cld2
is similar to that of the single-cloud case. In contrast, the average
density of cld2 is lower (and thus there is less mass above threshold)
in simulations \emph{w4a45} and \emph{w4a60}. The cores also fragment
in these cases. In these simulations cld2 is noteable for being in the
``middle'' of the cloud distributions. Fig.~\ref{fig:mor_s3w4} shows
that when cld2 is ``outside'' it is longer and narrower, whereas when
it is in the ``middle'' it is wider and shorter. 

Fig.~\ref{fig:core_mass} shows that the average core mass of cld3
at late times is similar to or slightly higher than that of an
isolated cloud (note that the symmetry of simulation \emph{w4a60}
means that cld3 behaves identically to cld1, while the symmetry of
simulation \emph{w4a0} means that cld3 is identical to cld2). Very
little fragmentation is seen in cld3 in any of the simulations, and in
particular in simulation \emph{w4a0} where cld2 is directly
alongside it.  In general the further downstream cld3 is, the more
mass is contained in the core, though this variation is quite small
and is somewhat time-dependent.

Fig.~\ref{fig:core_density} shows the evolution of the density in cld2
in three of the 3-cloud simulations. We see that as various shocks
pass through cld2 (the transmitted shock is the main one, but shocks
also propagate inwards from the sides and back of the cloud), the
average density increases by a factor of $~3-4$. Reexpansion starts
after $t \approx 1 \, t_{\rm cc}$ and the density drops reaching a
local minimum at $t \approx 2 \, t_{\rm cc}$. The density then
increases slightly due to compression from the ram pressure of the
flow as the cloud is accelerated downstream. The density steadily
decreases from $t \approx 3 \, t_{\rm cc}$ as the acceleration subsides
and as material is stripped away. In simulation \emph{s3w4a30}, cld2
lies in the ``flux rope'' of cld1 and is largely shielded from the
flow. As a consequence it does not experience a period of
re-compression at $t\approx 3 \, t_{\rm cc}$, but neither does it
experience strong stripping by the flow. At $t \approx 4 \, t_{\rm
  cc}$, cld1 collides with cld2 and the density of cld2 steadily
increases up to $t = 9\, t_{\rm cc}$.

Fig.~\ref{fig:s3_r128} shows that the evolution of $\beta$ in the
material of cld1 and cld3 is largely independent of the cloud
arrangement. However, this is not the case for cld2, where clear
differences can be seen between simulations in the second and third
panels of Fig.~\ref{fig:s3_r128}. However, this is hardly surprising,
since cld2 is variously located in the ``flux rope'' of cld1 in
simulation \emph{w4a30}, in the ``wings'' of cld1 in simulations
\emph{w4a15} and \emph{w4a45}, in the ``outside'' flow in simulation
\emph{w4a0}, and in the ``inside'' flow in simulation
\emph{w4a60}. The presence of a third cloud appears to modify the
behaviour seen in Fig.~\ref{fig:s2o8} - specifically $\beta$ is higher
when cld2 is between cld1 and cld3 (as in simulations
\emph{w4a45} and \emph{w4a60}).

\begin{figure}
  \begin{center}
    \includegraphics[width=7cm]{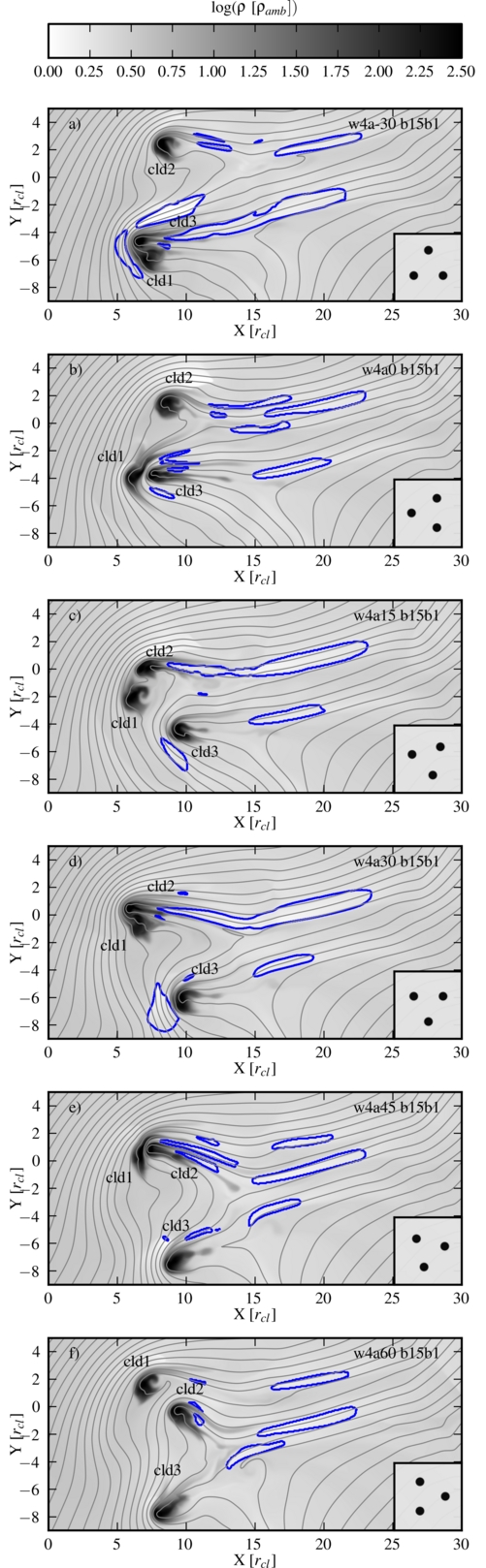} \\
    \caption{As Fig.~\ref{fig:mor_s3w4} but for an oblique shock
      ($\theta_0 = 15^\circ$, $\beta_0=1.13$). All snapshots are at
      $t= 4\,t_{\rm cc}$.}
    \label{fig:mor_s3w4_ob}
    \end{center}
  \end{figure}

%\afterpage{\clearpage} %Flush floats

\begin{figure}
  \begin{center}
    \includegraphics[width=7cm]{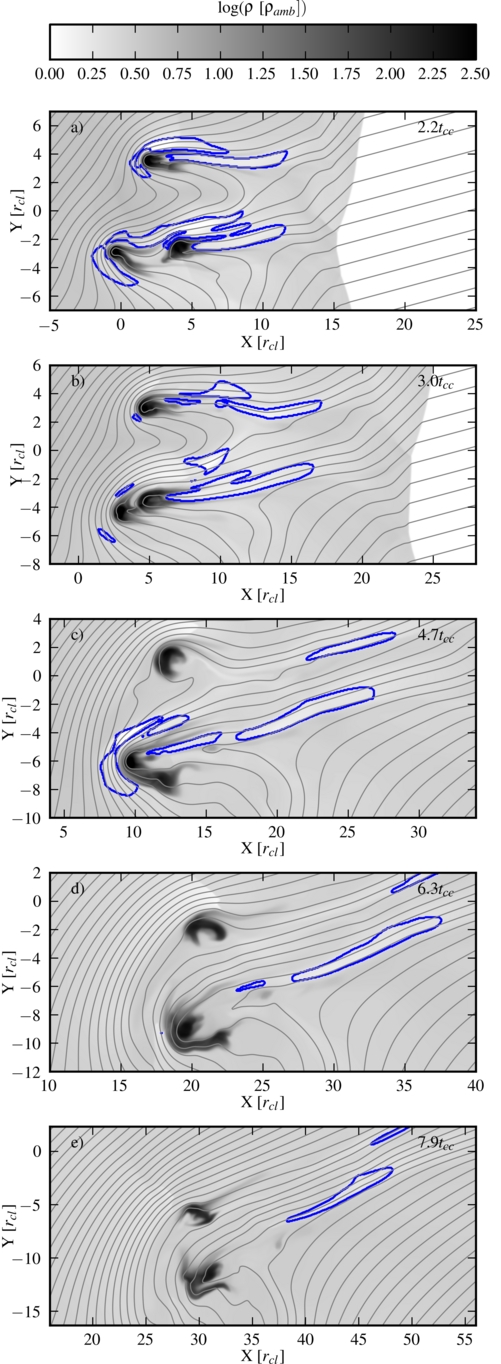} \\
    \caption{The time evolution of an oblique shock
      ($\theta_0=15^\circ$, $\beta_0 = 1.13$) interacting with
      3-clouds (simulation \emph{s3w4a-30}, $a = -30^\circ$).}
    \label{fig:evo1_s3w4_ob}
    \end{center}
  \end{figure}

\begin{figure}
  \begin{center}
    \includegraphics[width=7cm]{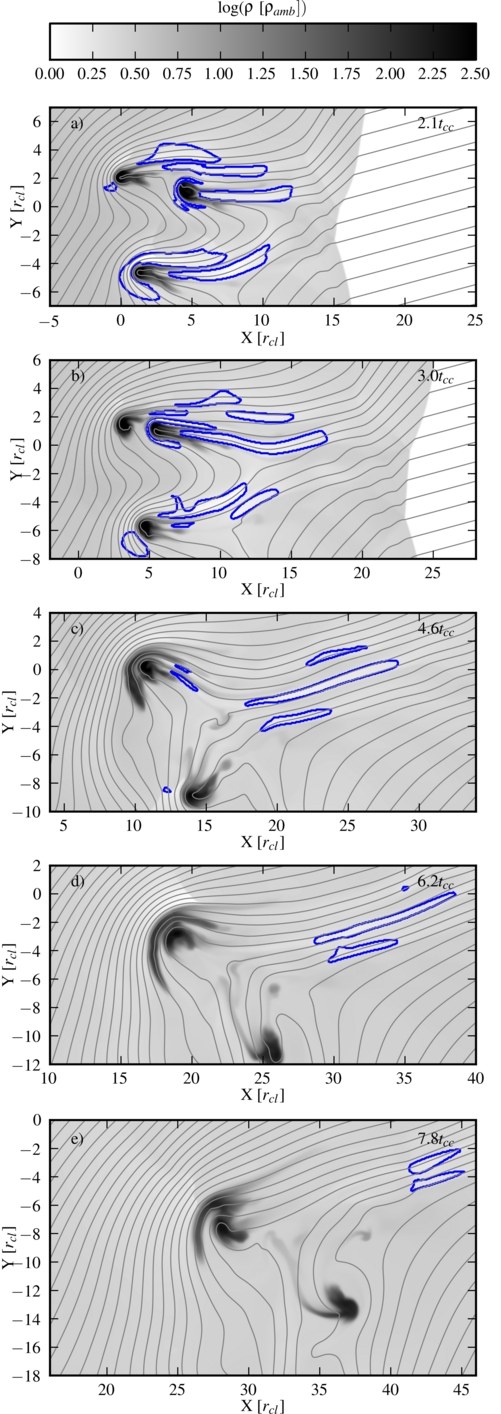} \\
    \caption{As Fig.~\ref{fig:evo1_s3w4_ob} but for simulation
      \emph{s3w4a45} ($\theta_0=15^\circ$, $\beta_0 = 1.13$, $a = 45^\circ$).} 
    \label{fig:evo2_s3w4_ob}
    \end{center}
  \end{figure}

\begin{figure}
  \begin{center}
    \begin{tabular}{c}
       \includegraphics[width=7cm]{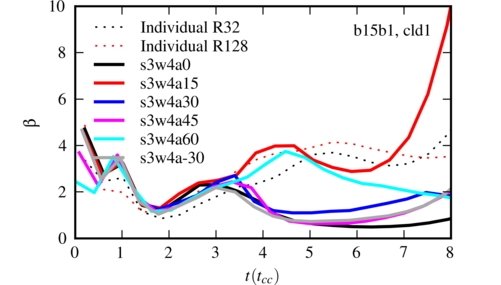} \\
      \includegraphics[width=7cm]{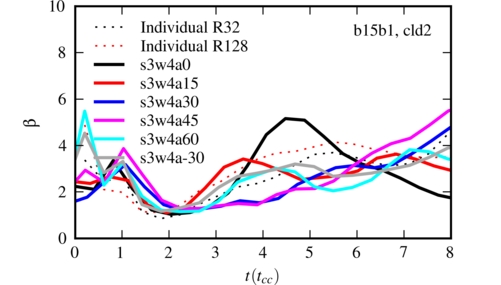} \\
      \includegraphics[width=7cm]{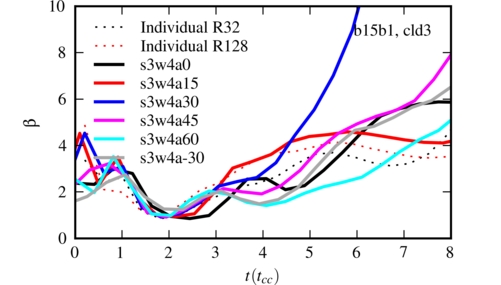} \\
    \end{tabular}
    \caption{The evolution of the harmonic mean of $\beta$ for 3-cloud
      simulations with an oblique magnetic field. The top, middle, and
      bottom panels show $\beta$ for cld1, cld2 and cld3
      respectively. The time axis is shifted appropriately for each
      cloud. The evolution of $\beta$ in isolated clouds is also shown
      (for simulations with 32 ($R_{32}$) and 128 ($R_{128}$) cells
      per cloud radius).
  }
    \label{fig:s3c}
  \end{center}
\end{figure}

\subsubsection{Oblique shocks}
\label{sec:3clouds_oblique}
We now study the interaction of 3-cloud distributions with an oblique
shock ($\theta_{0}=15^{\circ}$).  Fig.~\ref{fig:mor_s3w4_ob} shows the
resulting morphology at $t = 4\,t_{\rm cc}$. An additional simulation
with a negative orientation angle is also included (simulation
\emph{w4a-30}). In the \emph{w4a-30} and \emph{w4a0} simulations, the
modifier cloud is cld2\footnote{Naively we expect the switch to happen
  at an angle $a \approx 5^\circ$.}, but otherwise it is cld3.  A two
stage process occurs: firstly, cld1 interacts (as in the 2-cloud case)
with the nearest cloud along the flow, then these clouds jointly
interact with the third cloud. For instance, simulation \emph{w4a-30}
in Fig.~\ref{fig:mor_s3w4_ob}a) can be deconstructed as cld1 and cld3
interacting as in simulation \emph{w0o8} in Fig.~\ref{fig:mor_o8_ob},
and then the resulting combined ``clump'' interacting with cld2 as in
simulation \emph{w4o-4} in Fig.~\ref{fig:mor_w4_ob}. Similarly,
simulation \emph{w4a60} in Fig.~\ref{fig:mor_s3w4_ob}f) shows cld1 and
cld2 interacting as in simulation \emph{w4o4}, and then together
interacting with cld3 as in simulation \emph{w4o0} (compare
Fig.~\ref{fig:mor_s3w4_ob}a with Fig.~\ref{fig:mor_w4_ob}d).  The
secondary interaction can also be categorised in terms of a ``width''
and an ``offset''. In the 3-cloud simulations studied, it appears that
the appropriate width is the average ``width'' between the combined
clump and the third cloud, while the appropriate offset is between the
more upstream of the two clouds interacting in the first stage and the
third cloud with which they interact in the second stage\footnote{So
  it is possible to make a-priori estimates of these values.}.  Note
that the secondary interaction has a greater effective ``width'' than
the 2-cloud cases considered in Sec.~\ref{sec:2cloud}. This means that
the separation at closest approach is greater and that a secondary
collision between the combined clump and the third cloud does not
occur. However, otherwise the morphologies are roughly equivalent.

Fig.~\ref{fig:evo1_s3w4_ob} shows the time evolution of simulation
\emph{s3w4a-30} while Fig.~\ref{fig:evo2_s3w4_ob} shows the time
evolution of simulation \emph{s3w4a45}. In simulation \emph{s3w4a-30},
cld1 is initially at the bottom-left of the distribution, cld2 is at
the top-right, and cld3 is at the bottom right (see also
Fig.~\ref{fig:mor_s3w4_ob}a). As the shock sweeps over, cld1 moves
towards cld3 which is in the lee of cld1. cld1 engulfs cld3 by
$t\sim4\,t_{\rm cc}$, and cld3 is then confined by the magnetic field
threaded through cld1. In contrast, cld2 evolves in a relatively
isolated way. The flow tries to force its way between cld1/3 and cld2,
but the field lines between these two regions prevent this.  In
contrast, in simulation \emph{s3w4a45} cld1 is intitially at the
top-left of the distribution, cld2 is the most downstream cloud, and
cld3 is at the bottom left (see also
Fig.~\ref{fig:mor_s3w4_ob}e). Fig.~\ref{fig:evo1_s3w4_ob} shows that
cld1 and cld2 interact first, and that cld1 engulfs cld2. Although
cld3 is initially upstream of cld2, cld3 lies downfield. Thus as the
interaction proceeds, the tension in the field lines created by the
flow causes cld3 to accelerate downstream faster than the other
clouds.

In the oblique field case cld1 often has very low $\beta$ at late
times (see Fig.~\ref{fig:s3c}). Low $\beta$'s at late times were
previously seen in the top cloud of the 2-cloud simulations in
Sec.~\ref{sec:2cloud} (see simulations \emph{w4o8}, \emph{w2o8} and
\emph{w0o8} in Fig.~\ref{fig:hav_s2o8_obl}). In each case this is
caused by the collision of the cloud with a cloud further downstream.
Fig.~\ref{fig:mor_s3w4_ob} reveals that in the two cases where $\beta$
stays higher (simulations \emph{w4a15} and \emph{w4a60}), cld1 has not
collided with another cloud by $t = 4 \, t_{\rm cc}$. In simulation
\emph{w4a15}, Fig.~\ref{fig:mor_s3w4_ob} shows cld1 about to squeeze
between the two other clouds. cld1 proceeds to move into the
``shadow'' of cld2, and $\beta$ in cld1 rapidly grows after $t = 6.5
\, t_{cc}$. In simulation \emph{w4a60}, cld1 and cld2 accelerate at a
similar rate and do not collide (Fig.~\ref{fig:mor_s3w4_ob} shows
these clouds still with significant separation at $t = 4\,t_{\rm
  cc}$). However, after $t = 6 \, t_{\rm cc}$, as these clouds get
close, $\beta$ decreases in cld1.

The evolution of $\beta$ in the other two clouds does not deviate much
from the single-cloud case (see the middle and bottom panels of
Fig.~\ref{fig:s3c}). The only noteworthy behaviour is that cld2
generally has a slightly lower $\beta$, while cld3 has a slightly
higher $\beta$, at late times. $\beta$ in cld2 is most different from
the single-cloud case for simulation \emph{w4a0} ($\beta$ becomes very
low by $t \gtsimm 7\,t_{\rm cc}$), while for cld3 it is simulation
\emph{w4a30} ($\beta$ becomes very large at $t \gtsimm 5\,t_{\rm
  cc}$).

\begin{figure}
  \begin{center}
    \includegraphics[width=7cm]{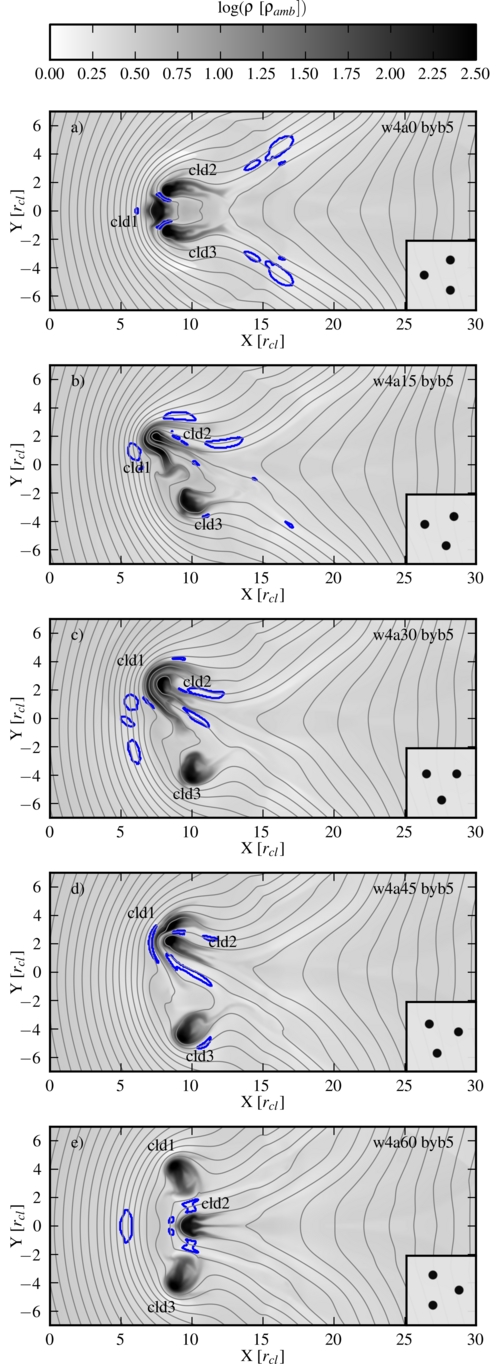} \\
    \caption{As Fig.~\ref{fig:mor_s3w4} but for a perpendicular
    shock ($\beta_0=5.06$). All snapshots are at $t = 4\,t{\rm cc}$.}
    \label{fig:mor_s3w4_by}
    \end{center}
  \end{figure}

\begin{figure}
  \begin{center}
    \includegraphics[width=7cm]{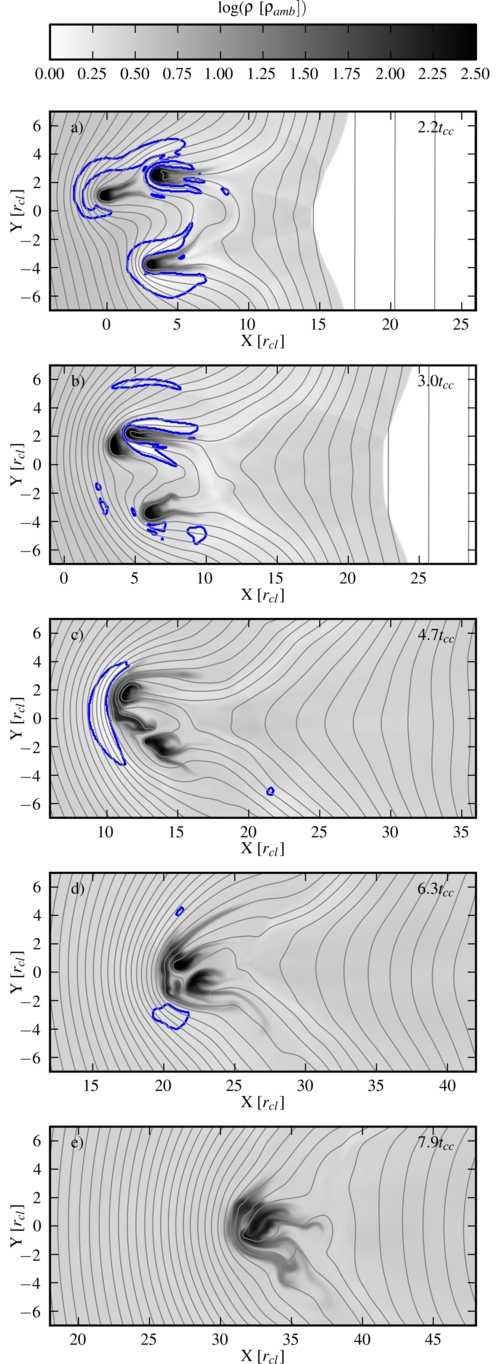} \\
    \caption{The time evolution of a perpendicular shock interacting
      with 3 clouds with $\beta_0 = 5.06$ (simulation \emph{s3w4a15}).}
    \label{fig:evo_s3w4_by}
    \end{center}
  \end{figure}

\begin{figure}
  \begin{center}
    \begin{tabular}{c}
      \includegraphics[width=7cm]{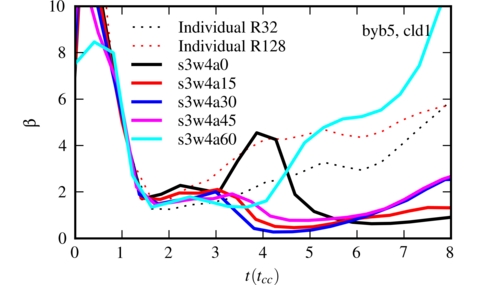} \\
      \includegraphics[width=7cm]{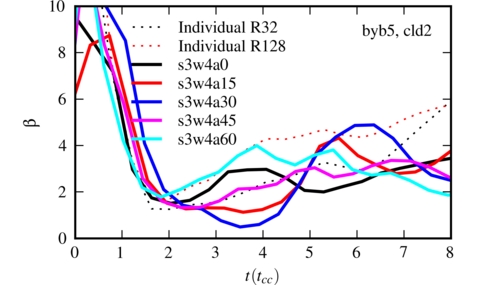} \\
      \includegraphics[width=7cm]{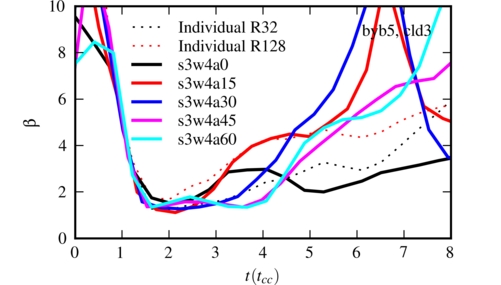} \\
    \end{tabular}
    \caption{The evolution of the harmonic mean of $\beta$ for 3-cloud
      simulations with a perpendicular magnetic field. The top,
      middle, and bottom panels show $\beta$ for cld1, cld2 and cld3
      respectively. The time axis is shifted appropriately for each
      cloud. The evolution of $\beta$ in isolated clouds is also shown
      (for simulations with 32 ($R_{32}$) and 128 ($R_{128}$) cells
      per cloud radius). }
    \label{fig:s3c_beta_perp}
  \end{center}
\end{figure}

\subsubsection{Perpendicular shocks}
\label{sec:3clouds_perpendicular}
In this section we study the interaction of a perpendicular shock with
3 closely spaced cylindrical clouds. Fig.~\ref{fig:mor_s3w4_by}
illustrates the range of morphologies which exist at $t = 4\,t_{\rm
  cc}$ from a variety of simulations. It reveals that collisions are
common. The collisions increase the density of the downstream cloud of
the pair and in some cases can last up to $t \sim 10 \, t_{cc}$ (cf.
Fig.~\ref{fig:evo_s3w4_by}). In all cases the magnetic field in the
oncoming flow is unable to pass between the clouds. It instead piles
up at the upstream side and the field lines then bend around the
clumpy region. Clouds either side of the center of the region then
behave like the ``top'' cloud in the 2-cloud oblique simulations
(cf. Sec.~\ref{sec:2cloud_oblique_morphology}).

Fig.~\ref{fig:evo_s3w4_by} shows the time evolution of simulation
\emph{s3w4a15}. cld1 is initially accelerated towards cld2 and cld3,
and at $t = 4.6\,t_{\rm cc}$ it appears to be poised to squeeze between
them. However, the snapshot at $t = 6.3\,t_{\rm cc}$ reveals that this
does not happen. Instead, the field line that cld1 sits on is not able
to force its way between cld2 and cld3, and cld1 ends up spreading
along it while the field line instead wraps around cld2 and cld3. At
the same time, cld2 and cld3 are forced together and mostly merge
(they are on similar field lines). The level of mixing depends on the
field strength and the degree of diffusion of material across the
field lines. The field lines straighten out at later times as the
clouds are accelerated up to the flow speed of the post-shock gas. It
is clear that the overall ``x''-size of the clumpy region is reduced
by the field compression in this direction, while the ``y''-size is
reduced by the diffusion of clouds along the field lines.

Fig.~\ref{fig:s3c_beta_perp} shows the evolution of $\beta$ in the
material of cld1, cld2, and cld3 in simulations with a perpendicular
field ($\beta_0=5.06$). In general, we see that $\beta$ in cld1 is
much lower than the isoated single cloud case, except for simulation
\emph{s3w4a60}. This simulation is noteable because it is the only one
in which cld1 is sufficiently on the ``outside'' of the distribution
that it does not collide with any of the other clouds (see
Fig.~\ref{fig:mor_s3w4_by}). Fig.~\ref{fig:s3c_beta_perp} also shows
that the $\beta$ in cld2 is similar to but generally lower than the
isolated cloud case. $\beta$ is most variable in simulation
\emph{s3w4a30} (in cld2 it is low at $t = 3.5-4\,t_{\rm cc}$ when cld1 is
compressing cld2, becomes noticeably higher at $t = 6\,t_{\rm cc}$,
and then drops again afterwards as it interacts strongly with cld3). The
value of $\beta$ in cld3 shows the most difference between
simulations. For \emph{s3w4a0} it stays low for most of the simulation
time, but for simulations \emph{s3w4a15} and \emph{s3w4a30} $\beta$
becomes very high at $t \approx 6.5\,t_{\rm
  cc}$. Fig.~\ref{fig:evo_s3w4_by} shows that in simulation
\emph{s3w4a15}, cld3 moves into the lee of cld1 at about this time (so
is sheltered), but by $t=7.9\,t_{\rm cc}$ cld1 has collided with it,
decreasing $\beta$ once more.

\section{Summary and conclusions}
\label{sec:conclusions}
The results shown in Sec.~\ref{sec:results} illustrate that the
presence of nearby clouds modifies the evolution of a shocked
cloud. In general, clouds on the same field lines are able to merge,
even if they are quite widely separated. Conversely, clouds on
different field lines tend to ``rebound'' from each other if they are
squeezed closely together. However, the details of the simulations are
complicated. We now summarize the main results and attempt to draw
generalities where possible, commenting on parallel, oblique and
perpendicular shock interactions in turn.

In the case of a parallel shock, the shocked cloud needs to push aside
fieldlines in order to expand laterally and this is made more
difficult by a cloud alongside. Hence the expansion and fragmentation
of the cloud is reduced. The downstream cloud is not very sensitive to
the distance along the direction of the shock normal to the upstream
cloud, at least for the range studied (``offsets''$~1-8 \,
r_{cl}$). Rather, for parallel shocks, the separation of clouds
perpendicular to the shock normal (i.e. their ``width'') largely
determines their evolution. As the field lines disturbed by the
upstream cloud advect downstream, they curl round and confine any
downstream cloud separated by ``widths''$~1-4 \, r_{cl}$. At
``widths'' of $~4 \, r_{cl}$ the evolution of clouds is analogous to
the evolution of clouds alongside one another (i.e. with an
``offset''$\approx 0$).
At a ``width'' of $~2 \, r_{cl}$, the downstream cloud is confined and
roughly circular, with mass stripping occuring along a tail from its
outside edge. Such clouds are pushed towards the lower pressure region
behind the upstream cloud and start expanding once in the lee. At
negligible ``widths'' a downstream cloud can fall in the ``flux-rope''
of the upstream cloud. While the initial shock compression of the
downstream cloud is comparable to that of an isolated cloud, it is
subsequently shielded from the flow and is not compressed nor
accelerated significantly. After shock compression and re-expansion
the properties of the downstream cloud are relatively constant until
the upstream cloud ploughs into it (i.e. the evolution of a cloud in a
flux rope is delayed until the upstream cloud reaches it).

In general, the presence of clouds downstream increases $\beta$ in the
upstream cloud via mechanical interaction, while clouds alongside
decrease $\beta$ by suppressing lateral expansion. By far the biggest
effect is when a cloud is directly behind and in the ``flux-rope'' of
an upstream cloud: in this case $\beta$ in the downstream cloud can be
significantly reduced for an extended period of time.

This basic behaviour also holds when a parallel shock interacts with
three clouds, though the additional cloud modifies the morphology
slightly. The additional cloud now allows a distinction to be made
concerning whether the downstream cloud lies ``inside'' or ``outside''
with respect to the rest of the distribution (e.g., simulation
\emph{w4a15} vs. simulation \emph{w4a45}). An outside cloud is
confined much as in the 2-cloud simulations, but the field lines
cannot curl as much around an inside cloud. The plasma $\beta$ is
generally higher in inside clouds, yet they are less confined than
outside clouds.

The interaction of an oblique shock with clouds is a more general case
than the specific cases of interactions of parallel or perpendicular
shocks. With oblique shocks, as well as considering whether a cloud is
upstream or downstream, one must also consider whether it is upfield
or downfield. In 2-cloud interactions we see some interesting dynamics
where the upstream cloud accelerates past the downstream cloud, and
then swings into its lee. The ``shielded'' cloud then experiences
reduced confinement forces and begins to diffuse, while the cloud more
exposed to the oncoming flow experiences another period of
compression. Clouds are given much faster transverse motions than
those interacting with parallel or perpendicular shocks.  The plasma
$\beta$ in the upstream cloud can drop below unity for a duration of a
few $t_{\rm cc}$ when it collides with the downstream cloud. The
interaction of an oblique shock with three clouds shows the same type
of behaviour, and can be understood in terms of the interaction of the
most upstream cloud with its nearest neighbour, and then their joint
interaction with the remaining cloud.

The interaction of a perpendicular shock with clouds is again a more
specific case. If the clouds are side-by-side they have a chance of
merging. We clearly see this in simulations where the clouds are
separated with an initial ``width'' of $4\,r_{\rm cl}$, but as the
width is increased the clouds should eventually evolve as isolated
clouds. We have not explored the transition between these regimes, but
it will certainly depend on paramters such as $M$, $\chi$ and
$\beta_{0}$. When the clouds have a non-zero ``offset'' the fact that
they exist on separate field lines prevents them from fully
mixing. Nevertheless, the clouds tend to be driven towards each other
much more strongly than when the shock is parallel or oblique. If the
clouds have a small ``width'' and larger ``offset'' the upstream cloud
tends to get driven into and then wraps around the downstream
cloud. Like the oblique case, the plasma $\beta$ in the upstream cloud
can become less than unity when it collides with a downstream cloud.
When three clouds are present, the most upstream or most downstream
cloud may be prevented from moving between the other two clouds due to
the tension in the field. Because the field lines also prevent the
flow from passing between the clouds the magnetic field builds up on
the upstream side and then bend around the clumpy region.

Previous work examining the MHD interaction of a shock with a single
cloud found that the plasma $\beta$ is low where the flow is
compressed, rather than the magnetic field being turbulently
amplified. The 2-cloud and 3-cloud interactions presented in this work
are more turbulent than single-cloud interactions due to the presence
of neighbouring clouds, but low values of $\beta$ are still not seen
very often. When they are, it is again mostly due to the compression
of the field by the flow, and is ultimately transient in nature.  This
highlights the difficulty of obtaining regions of low $\beta$ (e.g.,
$\beta < 1$) in adiabatic simulations. To obtain such regions it is
probably necessary to invoke cooling to reduce the thermal pressure
\citep[e.g.,][]{2007A&A...471..213V,2010MNRAS.406.1260V}.
cite{JohanssonZiegler2013} find that a weak perpendicular
field ($\beta \sim 10^3$) is able to suppress conduction without
limiting compression resulting in the highest density compressions of an
individual cloud. Without considering the cooling, we find that
moderate fields ($\beta=5$) are effective at bringing several clouds
together.

We note that the interaction of magnetized clouds has
  also been studied in solar physics, where \citealt{Shen2012} modelled the
  propagation and collision of two coronal mass ejections (CMEs) in
  interplanetary space. The resulting structures and their evolution
  resemble some of the work shown in the present paper, though it is
  clear that additional complexities, such as magnetic reconnection in
the neighbourhood of boundary layers (c.f. \citealt{Chian2012},
occur. Reconnection in turbulent flows is discussed in \citealt{Lazarian2014}). 

We now offer some thoughts on some important questions concerning
  the ISM. At this stage it is difficult to say anything about diffuse
  cloud lifetimes because the clouds in the simulation are 2D instead
  of 3D and some important physical processes, such as cooling and
  conduction, were not included. However, it is clear that the
  lifetimes are affected by the environment around the cloud, and
  specifically the presence of nearby clouds which can affect the flow
  and field lines. We have not considered specific observables in this
  work (such as emission maps), so it is unclear what types of
  structures would actually be visible. We note that some other works
  which have focussed on observables have considered high velocity
  clouds \citep{2012ApJ...751..120S,2012ApJ...753...58H}, supernova
  remnants \citep[e.g.,][]{2005ApJ...633..240P,2014MNRAS.437..976O}, and galactic winds
  \citep[e.g.,][]{2005MNRAS.362..626M}.  These works indicate that it
  is possible to gain some insights into some of the key parameters,
  such as the interstellar magnetic field, the Mach number of the
  shock, the properties of the clumpy medium, and the nature of the pressure sources. Insight into such
  parameters is most forthcoming, of course, when specific sources are
  modelled.

The present study has illustrated some of the complexity inherent in
MHD interactions of a shock with multiple clouds, and attempts to lay
some of the necessary foundations for understanding this problem.  In
future work we will build on the present study to examine the MHD
interaction of a shock with many tens and hundreds of clouds. We will
also extend this work to spherical as opposed to cylindrical
clouds. The interaction could be quite different between these two
cases because field lines will be able to slip past spherical
clouds,  which could significantly change the forces acting on the
  clouds. In addition, there could be interesting interactions between
  clouds whose field lines lie in different planes. For instance,
  consider the interaction of a cloud in one plane with a second cloud
  in an adjacent parallel plane where there are different field lines
  in each plane. If the planes are far enough apart then the clouds
  should evolve independently (one plane might slip sideways relative
  to the other). However, the evolution may be markedly different when
  the planes are close enough together that pressure interactions
  occur between them.

\section*{acknowledgements}
We would like to thank the referee for a helpful report which improved
this paper. JMP would like to thank the Royal Society for previously funding
a University Research Fellowship and STFC for continued support. 

\bibliography{biblio}

\end{document}